\documentclass[twocolumn]{aastex631}
\usepackage{amssymb}
\usepackage{amsmath}
\usepackage{appendix}

\newcommand{\erg}{\text{~erg}}
\newcommand{\ergs}{\text{~erg}\text{~s}^{-1}}
\newcommand{\cm}{\text{~cm}}
\newcommand{\s}{\text{~s}}



\shortauthors{Chen $\&$ Dai}
\shorttitle{Jets in AGN}

\begin{document}
	
	\title{Observational Properties of Nonthermal Emission from Relativistic Jets Escaping Active Galactic Nucleus Disks}
	
	\author[0000-0001-8955-0452]{Ken Chen}
	
	\author[0000-0002-7835-8585]{Zi-Gao Dai}
	
	\affiliation{Department of Astronomy, School of Physical Sciences, 
		University of Science and Technology of China, Hefei 230026, People’s Republic of China; cken@ustc.edu.cn, daizg@ustc.edu.cn}
	
\begin{abstract}
Relativistic jets launched from stellar-mass compact objects embedded in the 
accretion disk of an active galactic nucleus (AGN) can produce nonthermal 
emission upon successfully breaking out of the disk. 
In this paper, we present a comprehensive study of the long-term propagation 
dynamics and broadband nonthermal radiation signatures of such jets in a 
realistic AGN environment, explicitly modeled 
as wind outflows. Our modeling reveals 
two distinct features imprinted by the high-density AGN medium: rapid 
deceleration of the jet ejecta, accompanied by a prompt downshift of the 
emission spectral energy distribution, and persistently strong synchrotron 
self-absorption, giving rise to a prominent quasi-thermal hump in the 
emission spectrum. Crucially, both gamma-ray burst jets and jets powered by 
accreting binary black hole merger remnants can produce detectable 
multi-wavelength emissions that substantially outshine the AGN background. 
Moreover, the short time delays between gravitational wave triggers and 
these electromagnetic counterparts\textemdash typically less than 
$10^6\s$\textemdash greatly facilitate secure multi-messenger associations. 
Besides, our findings highlight that interaction-induced radiation from 
AGN-embedded jet systems offers a powerful diagnostic probe of the 
spatial distribution, density structure, and physical properties of the AGN 
medium.
\end{abstract}

\keywords{Active galactic nuclei (16); Relativistic jets (1390); Gravitational wave sources (677); nonthermal radiation sources (1119);}

\section{Introduction}
Relativistic jets with diverse powers and durations can be launched from
a variety of extreme astrophysical events occurring in the accretion
disk of an active galactic nucleus (AGN) \citep{Chen25}. In the era of 
multi-messenger astronomy, two representative classes of such disk-embedded 
jets have drawn particular attention: those associated with gamma‑ray burst
(GRB) and those driven by binary black hole (BBH) mergers.

GRBs originate either from the core collapse of massive stars 
\citep{Woosley93, Woosley06a} or from mergers of binary neutron star (BNS) 
and neutron star-black hole (NSBH) systems \citep{Abbott17a,Abbott17b}. 
Stars embedded in AGN disks can attain high masses and rapid rotation 
via efficient accretion \citep[e.g.,][]{Dittmann21,Jermyn21,Fabj25}, 
rendering them promising progenitors for long GRBs \citep{Woosley06b}. 
AGN disks are also considered fertile environments for BNS and NSBH mergers 
\citep{McKernan20, Yang20, Perna21a}. Motivated by these prospects, 
extensive efforts have been devoted to modeling GRB jets within AGN disks,
encompassing jet propagation \citep{Zhu21, Zhang24}, the production and
diffusion of prompt emission and afterglow radiation 
\citep{Perna21b, Lazzati22, Wang22, Ray23, Kang25, Zhao26}, as well as the
roles of jet structure \citep{Kathirgamaraju24, Yuan25}, time-dependent
energy injection \citep{Huang24}, and prolonged central-engine activity 
\citep{Wei25}. Nevertheless, all existing studies confine their
environmental modeling exclusively to the AGN disk. 
Even when jet dynamics and afterglow evolution
after disk breakout are considered, the overlying ambient medium is
routinely approximated as a uniform interstellar medium (ISM), which is 
a substantial oversimplification.

AGN disks are also compelling sites for BBH mergers 
\citep[e.g.,][]{Tagawa20,Ishibashi24,McKernan25,Rowan25,
Delfavero25,Ford25,Li25, Zhu26}, 
potentially accounting for anomalous gravitational‑wave(GW) events such as 
GW190521 \citep{Abbott20,Graham20} and 
GW23123 \citep{Abac25, He25, Bartos26, Li26}. Moreover, 
the gas‑rich environment of an AGN disk enables the post-merger remnant 
to launch a relativistic jet through 
accretion \citep[e.g.,][]{McPike26}. Thermal emission 
produced during jet breakout from the disk and the subsequent expansion of 
the cocoon formed by jet-AGN disk interaction have been investigated by 
several studies \citep{Tagawa23,Tagawa24,Tagawa26,Chen24,Chen25,Rodriguez-Ramirez24}. 
To date, however, only \citet{Wang21} has explored the dynamical evolution 
and consequent radiation following the jet ejecta emergence from the disk. 

AGN disks are geometrically thin, occupying only a small fraction of the 
overall AGN space. Thus, once a jet successfully breaks out of the disk, 
its subsequent interaction with the overlying AGN environment gives rise to
nonthermal emission that persists far longer than its thermal 
counterpart, rendering it a crucial observational probe of such systems. 
Moreover, the AGN environment above the disk is far from an ISM‑like state, 
and is instead characterized by high‑density gas—manifesting as clumpy clouds or 
large‑scale outflows—and an intense radiation field \citep{Netzer13}. 
Despite its importance, the long‑term propagation of jets in this 
environment and the resulting broadband nonthermal radiation remain largely 
unexplored.

\begin{figure*}
	\begin{center}
		\includegraphics[width=0.45\textwidth]{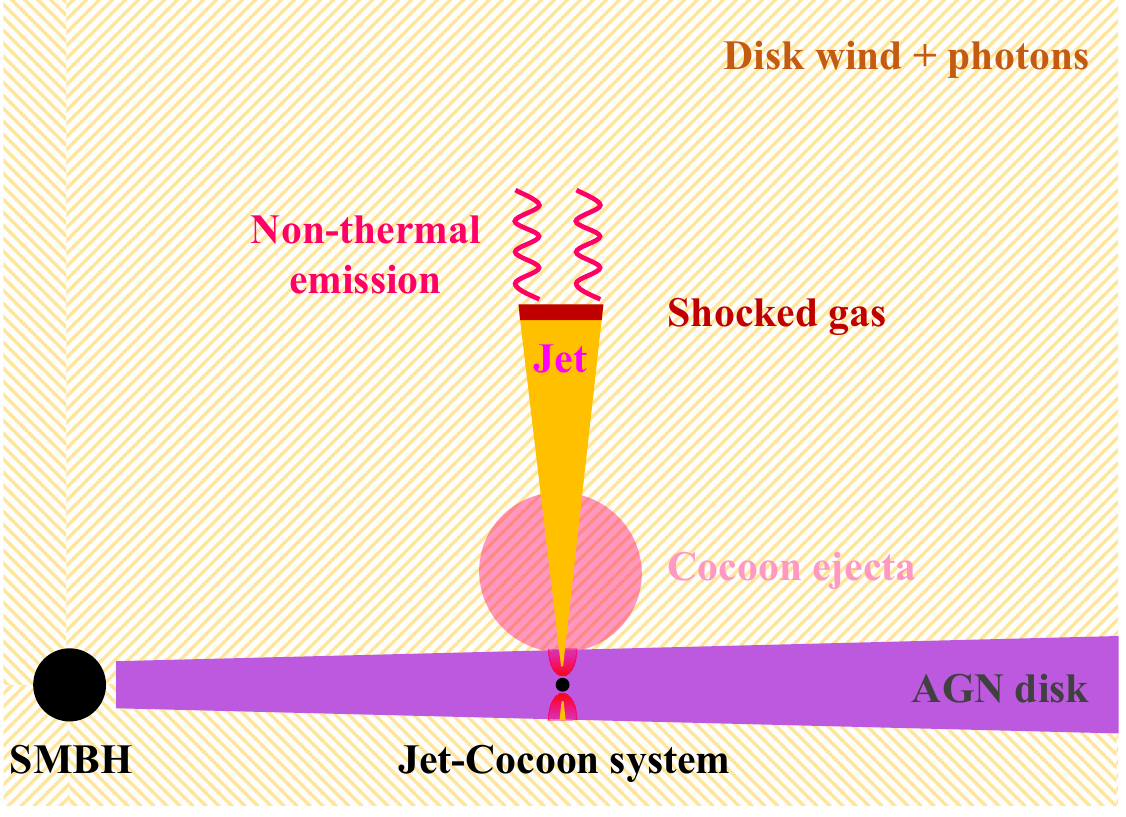}

	\end{center}
	\caption{Schematic description of a jet successfully
		breaking out of the AGN disk and propagating through 
		the AGN environment. AGN-driven wind outflows and 
		AGN photons fill the region above the disk. The
	    interaction between the jet and the ambient
	    medium drives strong shock, producing nonthermal
	    emission via particle acceleration. Besides, the 
	    cocoon, formed by the energy deposition from the 
	    jet propagating within the disk, expands under its 
	    own pressure.}	
	\label{Fig:sch}
\end{figure*}

In this paper, we investigate the observational signatures of nonthermal 
emission generated by a relativistic jet that successfully escapes the AGN
disk and subsequently propagates through the overlying AGN medium.
A schematic diagram is depicted in Figure \ref{Fig:sch}.
We model the co-evolution of propagation dynamics and multi-wavelength 
emission for jets with varying power and duration, explicitly accounting 
for their interaction with the high-density AGN gas. We further 
assess their detectability and calculate the resulting multiband light 
curves.  As a complementary study to \cite{Chen25}, this work completes 
the radiation characteristics of the AGN-disk jet beyond the thermal 
breakout phase. The paper is organized as 
follows. In Section \ref{Sec2}, we describe our modeling of the AGN gas 
distribution, the jet dynamics, and the emission processes. Section 
\ref{Sec3} presents a comprehensive analysis of the nonthermal emission for 
three representative jet systems—a powerful uncollimated jet, a
	long‑lived free jet, and an
under-accelerated jet at breakout—focusing specifically on GRB 
jets and BBH merger remnant‑driven jets. In Section \ref{Sec4}, we discuss 
the detectability and multi‑band light curves of these jets. We summarize 
our main results and provide discussion in Section \ref{Sec5}.  
Symbols $G$ and $c$ in this paper represent the gravitational constant and 
the speed of light, respectively.

\section{Jet propagation in AGN environment}
\label{Sec2}

\subsection{AGN gas distribution}
To investigate the long-term propagation 
of a jet initially embedded in an AGN 
accretion disk throughout the broader AGN 
environment, the gas distribution 
in both the AGN disk and the circum-disk 
medium is required. For the AGN disk, 
following our previous work \citep{Chen25}, 
we adopt the radial structure model of 
\cite{SG03}, where at each disk 
radius $R_0$, the density
$\rho_\text{d}(R_0)$ and scale height
$H_\text{d}(R_0)$ of the disk are 
explicitly specified. The disk is 
approximated as a thin slab vertically 
bounded by $H_\text{d}$, with a 
uniform vertical gas number density 
profile given by
\begin{equation}
n= \rho_\text{d}(R_0)/m_{\mathrm{p}}, \quad r<H_{\text{d}} 
\end{equation} 
where $r$ is the vertical distance from 
the disk midplane, and $m_{\mathrm{p}}$
is the proton mass.

Above the disk, the observed broadening 
of emission lines indicates the 
presence of a large number of dense 
clouds, spanning density from 
$10^{8-10}\cm^{-3}$ in the broad-line 
region to $10^{3-5}\cm^{-3}$ in the 
narrow-line region, confined by a relatively 
low-density hot intercloud medium \citep{Netzer13}.
Besides the clouds, multiple classes of outflows 
are detected in most AGNs from the observation
of blue-shifted absorption lines \citep{Laha21}, 
confirming the prevalence of an outflow in an AGN
\footnote{The non-detection of these absorption
lines would reflect a sporadic nature of the 
outflow, which exhibits a low gas column density
feature during observation, rendering the line
becoming too weak to detect \citep{King15}.}.
Accretion disk winds provide a physically 
motivated origin for AGN outflows, with launch 
and acceleration processes feasible in AGN 
scenarios \citep[e.g.,][]{Proga00, Risaliti10, Giustini19}.
Moreover, these winds would naturally give 
rise to the clouds via condensation 
processes \citep{Netzer13}.
In practice, the launch and acceleration
mechanisms differ among distinct classes 
of AGN outflows \citep{Laha21}; aiming to 
characterize the key features of  
jet–AGN medium interaction, the wind 
structure is simplified into a 
uniform global form \citep{King15}
\begin{equation}
\dot{M}_{\mathrm{w}} \simeq 4 \pi b n_\text{w} R^{2} m_{\mathrm{p}} v_\text{w},
\end{equation}
filling the AGN environment, 
where $\dot{M}_{\mathrm{w}}$ is the wind
mass-loss rate, $b$ is the covering 
factor, $n_\text{w}$ is the gas number
density at a radial distance $R$ from the
central super-massive black hole 
(SMBH), one has
\begin{equation}
\begin{split}
n_{\mathrm{w}}= &6.2\times 10^7 \cm^{-3}\left(\frac{b}{0.5}\right)^{-1}
\left(\frac{\dot{M}_{\mathrm{w}}}{0.1\dot{M}_{\mathrm{Edd}}}\right) \times \\
&\left(\frac{R}{10^3R_{g}}\right)^{-2}\left(\frac{v_\text{w}}{10^9\cm \s^{-1} }\right)^{-1},
\end{split}
\end{equation}
for a SMBH of mass $M=10^7 M_\odot$, which
is consistent in magnitude with the 
observation \citep{Laha21}. The 
corresponding vertical gas density 
profile above the AGN disk is 
modeled as \citep[e.g.,][]{Zhou23}
\footnote{Note that $n$ exhibits a sharp discontinuity 
at the disk surface $H_\text{d}$. In reality, the 
density profile transitions continuously—from the 
AGN accretion disk, through the local disk wind, to 
the global AGN wind. Fortunately, the disk density 
declines rapidly above $H_\text{d}$, following 
either a Gaussian distribution 
$n \propto \exp\left[-r^2/(2 H^2_\text{d})\right]$ or 
a polytropic profile $n \propto (1-r^2/(6 H^2_\text{d}))^3$ 
\citep[e.g.,][]{Kato08}; within a few $H_\text{d}$, 
$n$ decreases from $\rho_\text{d}/m_{\mathrm{p}}$ to $n_0$.
This decline slightly modifies the initial properties of 
the successful jet \citep{Chen25}, such as reducing its 
duration after the breakout. Moreover, the jet deceleration
radius where the reverse shock becomes relativistic is 
significantly larger than $H_\text{d}$, as confirmed by 
Equation \eqref{dea}. Consequently, the detailed vertical 
structure of the AGN disk has only a minor influence on the 
jet long-term nonthermal emission.}
\begin{equation}\label{n}
n= n_0\left(\frac{\sqrt{R_0^2+r^2}}{\sqrt{R_0^2+H^2_{\text{d}}}} \right)^{-2}
\simeq n_0 \left[1+\left(\frac{r}{R_0}\right)^2 \right]^{-1}, 
\quad r>H_{\text{d}}
\end{equation}
where $R_0$ is the radius of the 
AGN disk where the jet launches,
$n_0=n_{\text{w}}
(\sqrt{R_0^2+H^2_{\text{d}}})
\sim 10^8 \cm^{-3}$ for 
$M=10^7 M_\odot$ and $R_0=10^3R_{g}$.
When the jet propagation distance 
is much smaller than $R_0$,
the ambient density remains approximately 
constant, i.e., $n\sim n_0$; at a
large distance, the density approaches
a wind-like profile, i.e., $n \propto r^{-2}$.
In this work, we neglect the ambient 
cloud medium. Jet-cloud interactions 
are expected to produce additional 
emission signals \citep[e.g.,][]{Zhuang25},
which will be 
studied in a future work.

\subsection{Jet dynamics}
\label{dyn}
\begin{figure*}
	\begin{center}
		\includegraphics[width=0.45\textwidth]{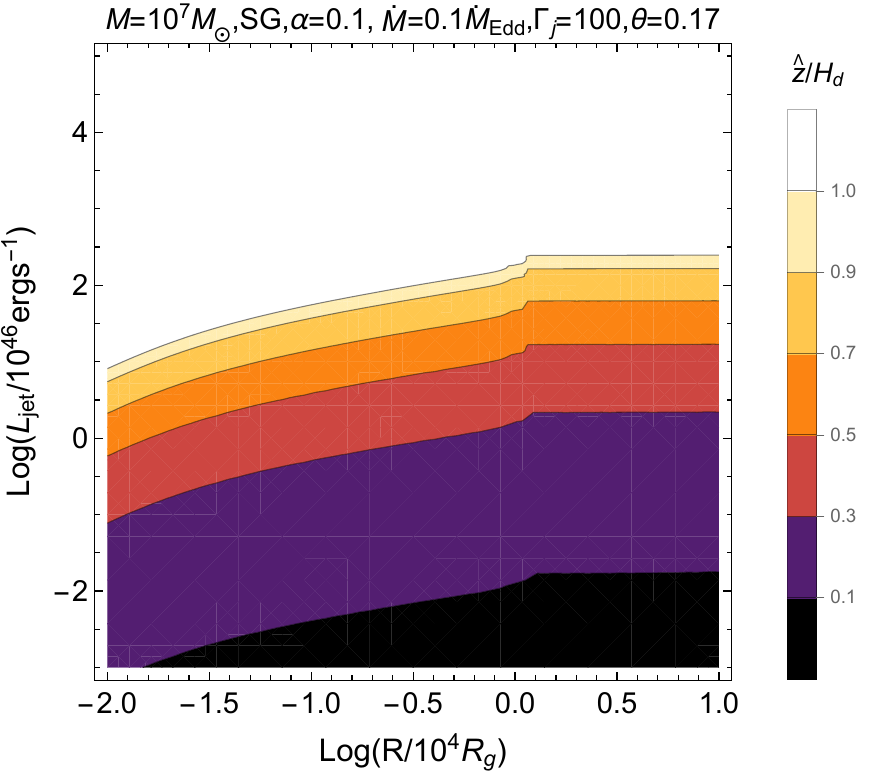}
		\quad
		\includegraphics[width=0.45\textwidth]{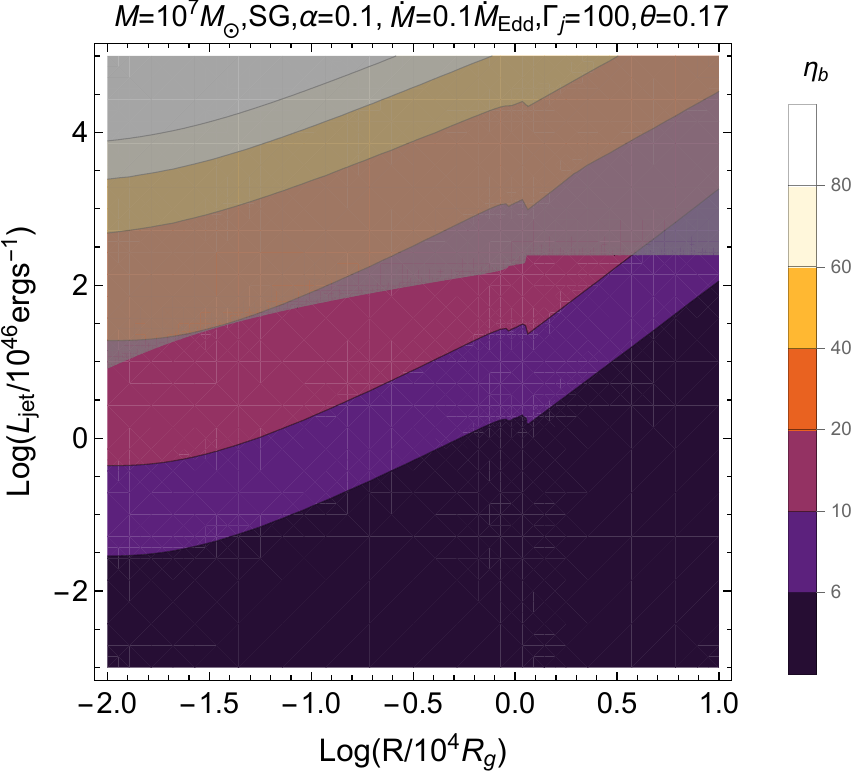}
	\end{center}
	\caption{Properties of the shocked materials in
	the successful jets at breakout. The left panel shows
	the distance from the jet base to the collimating 
	shock converge point, within which the jet remains
	predominantly unshocked, and beyond it the jet 
	material is shocked to resist the cocoon pressure.
    The right panel shows the critical baryonic loading
    of the shocked jet material, which determines the
    terminal LF the materials can be accelerated after
    breaking out of the AGN disk. The grey region 
    indicates parameter space where the entire jet 
    remains uncollimated and unshocked within the disk.}	
	\label{Fig:col}
\end{figure*}

When a jet propagates through an extremely dense 
environment, such as AGN disk, the interaction 
between its head and the surrounding medium continuously 
shocks gas, which then flows laterally to 
inflate a cocoon with large pressure, encasing and 
collimating the jet body 
\citep[e.g.,][]{Begelman89, Matzner03, Bromberg11}.
The dynamics of jet propagation within the AGN disk 
have been investigated in detail in \cite{Chen25}. 
As the jet head is significantly decelerated, for 
a jet with duration $t_{\text{j,cri}}$, the jet 
tail catches up to the head precisely at the time, 
$t_{\text{bre}}$, the head breaking out of the disk 
surface, resulting in the disappearance of jet body. 
Jets with duration $t_{\text{j,0}}< t_{\text{j,cri}}$
therefore become choked, where $t_{\text{j,cri}}$ 
denotes the critical jet duration, whose value is 
shown in Figure 5 of \cite{Chen25}. In this paper,
we focus on jets that successfully escape the AGN
disk, i.e., those with $t_{\text{j,0}}>t_{\text{j,cri}}$,
and for simplicity, assume the jet launches at the 
disk midplane, propagating vertically (perpendicular 
to the disk plane).  

For a successful jet, its power $L_\text{j}$ and 
duration $t_\text{j,0}$ jointly determine the 
initial properties when propagating in the AGN 
environment, specifically, the effective duration 
and the Lorentz factor (LF). On the one hand, 
as the leading segment of the jet with duration 
$t_{\text{j,cri}}$ transforms into the cocoon 
material, the remaining jet has a reduced duration
$t_\text{j}=t_\text{j,0}-t_{\text{j,cri}}$.
On the other hand, the overpressured cocoon 
compresses the jet body, driving a 
collimating shock inside the jet, of which
the convergence height is \citep{Bromberg11}
\begin{equation}
\hat{z} \simeq \left(\frac{L_{\text{j}}}{\pi c P_c}\right)^{\frac{1}{2}},
\end{equation} 
assuming that the jet acceleration radius is much 
smaller than $\hat{z}$, where the cocoon pressure 
$P_c$ is calculated following \cite{Chen25}. Below
$\hat{z}$, the jet propagates freely with an initial
LF $\Gamma_{\text{j}}=\Gamma_0$; above $\hat{z}$, 
the jet material passes through the collimating shock,
where it is shocked, thermalized, and decelerated to a
LF $\Gamma_{\text{j}}=1/\theta_0$, with most of its 
kinetic energy converted into internal energy, 
where $\theta_0$ is the initial opening angle of 
the jet \citep{Bromberg11}. 

As shown in Figure \ref{Fig:col},
when breaking out of the AGN disk at $t_\text{bre}$, 
a powerful jet remains uncollimated and largely 
unshocked, provided that $\hat{z}>H_{\text{d}}$ and 
the collimating shock has not yet converged.  
Conversely, if the collimating shock converges 
below the disk surface, the jet segment between 
$\hat{z}$ and $H_\text{d}$ is shocked prior to 
breakout. Consequently, the LF of the jet
at the onset of propagation in the AGN environment 
falls into three regimes. In the first regime, the 
remnant jet remains unshocked, with $\Gamma_j=\Gamma_0$, 
which requires $\hat{z}>H_{\text{d}}$. In the second 
regime, the jet duration satisfies
$t_{\text{j,cri}}<t_{\text{j},0}\lesssim t_\text{bre}$, so
the earliest-emerging jet segment is shocked and thermalized.
Subsequently, it undergoes self-acceleration and reaches 
a terminal LF of $\Gamma_{\text{j}}= \max[\Gamma_0, \eta_b]$, 
where $\eta_b$ denotes the critical baryonic 
loading, determining the maximum LF attainable before 
photon–baryon decouples and photon-acceleration 
halts, which is given by \citep{Nakar05, Nakar17}
\begin{equation}
\eta_b=\left(\frac{L_{\mathrm{j}} \sigma_{\mathrm{T}}}{2 \pi \theta_0^2 R_{\text{j}} m_{\mathrm{p}} c^{2}}\right)^{\frac{1}{4}},
\end{equation}
where $\sigma_\text{T}$ is the Thomson scattering 
cross-section, and $R_{\text{j}}$ is the lateral 
radius of the jet at its breakout. As shown in 
Figure \ref{Fig:col}, the shocked jet material
is inefficiently accelerated, forming a shell 
with a LF $\eta_b<\Gamma_0$. Subsequently, the 
trailing unshocked jet component catches up and 
collides with this slower shell, launching an 
internal shock that drives reacceleration. In the 
early phase following jet breakout, the expanding
cocoon maintains sufficient pressure to continuously 
compress jet segments embedded within the AGN disk, 
thereby producing additional shocked jet material, 
which persists for approximately a cocoon dynamical 
timescale $t_{\text{coc}}=R_{\text{c}}/v_{\text{c}}$
\citep{Mizuta13}. Calculating the jet-cocoon 
evolution within the AGN disk 
following \cite{Chen25}, we find that 
$t_{\text{coc}}\sim t_{\text{bre}}$. The third 
regime corresponds to a long-lived jet, defined 
by $t_{\text{j},0}\gg t_{\text{bre}}$, in which 
the cocoon thermal pressure declines continuously 
and dramatically during its expansion, rendering it 
incapable of compressing the jet. 
As a result, the bulk of jet material 
remains unshocked, and the jet retains its 
initial LF $\Gamma_j=\Gamma_0$.

After the jet is injected into the AGN environment
above the AGN disk and interacts 
with the swept ambient gas,
a forward shock (FS) is driven into the AGN medium, 
while a reverse shock (RS) simultaneously propagates 
back through the jet. Applying shock jump conditions 
and adopting the uniform pressure and LF  
assumption across the shocked AGN medium (region 2)
and the shocked jet (region 3), the LF
of the post-shock fluid can be calculated as 
\citep{Sari95, Metzger12}
\begin{equation}\label{g2}
\gamma_{2}=\gamma_{3}=\Gamma_{\mathrm{j}}\left[1+2 \Gamma_{\mathrm{j}}\left(\frac{n}{n_{\mathrm{j}}}\right)^{\frac{1}{2}}\right]^{-\frac{1}{2}},
\end{equation}
where 
\begin{equation}
n_{\mathrm{j}}= \frac{L_{\mathrm{j}}}{\pi 
	\theta^2 r^{2} m_{\mathrm{p}} c^{3} \Gamma_{\mathrm{j}}^{2}}
\end{equation}
is the jet density at the radius $r$, and 
$\theta$ is the jet opening angle. The
number density, internal energy density and
total number of protons in region 2 evolve as
\begin{gather}\label{eq:FS}
n_2 \simeq 4\gamma_2 n ,\\[1ex]
e_2 = 4(\gamma_2-1)\gamma_2 n m_p c^2,\\[1ex]
N_2 = \int_{H_{\text{d}}}^{r} \pi \theta^2 r^2 n dr.
\end{gather}

As the jet propagates through the dense AGN 
medium, the criterion for a RS to become 
relativistic, i.e. \citep{Zhang18}
\begin{equation}\label{dea}
\begin{split}
\Gamma_{\text{j}}^{2} / \frac{n_{\text{j}}}{n}=&137 \left(\frac{L_{\text{j}}}{10^{50}\text{~erg}\,\text{s}^{-1}} \right)^{-1}\left(\frac{\Gamma_{\text{j}}}{100}\right)^{4}\left(\frac{\theta}{0.1}\right)^{2} \times \\
&\left(\frac{n}{10^{8} \mathrm{~cm}^{-3}}\right)\left(\frac{r}{10^{15} \mathrm{~cm}}\right)^{2}\gg 1,
\end{split}
\end{equation}
is typically satisfied. 
Therefore, the parameters of relativistic RS evolve as 
\begin{gather}\label{eq:RS}
n_3 \simeq 4\gamma_{43} n_{\text{j}},\\[1ex]
e_3 = 4(\gamma_{43}-1)\gamma_{43} n_{\text{j}} m_p c^2,\\[1ex]
N_3 = \frac{L_{\text{j}}}{\Gamma_{\text{j}} m_p c^2} \frac{r}{2\gamma^2_3 c},
\end{gather}
where the relative LF between the jet shock
upstream and downstream is 
\begin{equation}
\gamma_{43} \simeq \frac{1}{2}\left(\frac{\Gamma_{\mathrm{j}}}{\gamma_3}+\frac{\gamma_3}{\Gamma_{\mathrm{j}}}\right).
\end{equation}
The radius at which the RS crosses
the entire jet is
\begin{equation}
r_{\text{c}}= \left(\frac{L_{\mathrm{j}} t^2_{\text{j}}}{\pi 
	\theta^2 n m_{\mathrm{p}} c}\right)^{\frac{1}{4}},
\end{equation}
and the corresponding observed 
crossing time is 
$t_{\text{cross}}=r_{\text{c}}/2\gamma^2_3 c$.
In this paper, an on-axis observer 
is considered, and viewing 
angle effects are not taken into account.

After the RS fully crosses the jet, 
the evolution of its downstream parameters 
is approximately described by 
\citep[e.g.,][]{Yi13}
\begin{equation}\label{RSS}
\begin{array}{l}\gamma_{3}=\gamma_{3,\mathrm{c}}\left(r / r_{\mathrm{c}}\right)^{\frac{(2 k-7)}{2}}, n_{3}=n_{3, \mathrm{c}}\left(r / r_{\mathrm{c}}\right)^{\frac{(2 k-13)}{2}}, \\[1ex] e_{3}=e_{3, \mathrm{c}}\left(r / r_{\mathrm{c}}\right)^{\frac{(4 k-26)}{3}}, N_{\mathrm{3}} = N_{3, \mathrm{c}} ,\end{array}
\end{equation}
where the subscript “c” denotes values evaluated
at $r_{\text{c}}$, and $k$ is an index
of the environmental density decay, which
is simply set to $0$ for $r<R_0$ and $2$ 
for $r>R_0$ based on the density distribution
of Equation \eqref{n}. $\gamma_{3}=1$ is
adopted as the termination condition for the 
RS evolution, since the relativistic self-similar 
solution described by Equation \eqref{RSS} becomes 
invalid once the RS transitions into the Newtonian 
regime. Although this threshold is artificial, it 
slightly influences the observed nonthermal emission
signature, because late-time emission is dominated 
by the FS.
 
For the FS, the fluid dynamics is governed
by the following coupled differential 
equations \citep{Huang00}
\begin{gather}\label{eq:dt}
\frac{d r}{d t}=\beta c \Gamma\left(\Gamma+\sqrt{\Gamma^{2}-1}\right),\\[1ex]
\frac{d m}{d r}=2 \pi r^{2}(1-\cos \theta) n m_{p},\\[1ex]
\frac{d \Gamma}{d m}=-\frac{\Gamma^{2}-1}{M_{\mathrm{j}}+\epsilon m+2(1-\epsilon) \Gamma m},\\[1ex]
\frac{d \theta}{d t} =\frac{c_{\mathrm{s}}\left(\Gamma+\sqrt{\Gamma^{2}-1}\right)}{r},
\end{gather}
where $\Gamma$, $m$, $\theta$ are the 
LF, mass, opening angle of the swept-up 
and shocked AGN medium, 
$M_{\text{j}}=L_{\text{j}} t_{\text{j}}/\Gamma_{\text{j}} c^2$
is the total jet mass, 
$\epsilon$ is the radiative efficiency of 
the fluid, i.e., the fraction of internal 
energy lost via radiation per dynamical 
timescale, and the sound speed is
\begin{equation}
c_{\mathrm{s}}^{2}=\hat{\gamma}(\hat{\gamma}-1)(\Gamma-1) \frac{1}{1+\hat{\gamma}(\Gamma-1)} c^{2},
\end{equation}
with the adiabatic index 
$\hat{\gamma} \sim (4\Gamma+1)/3\Gamma$.

\subsection{Emission equations}
\label{rad}
\begin{figure*}
	\begin{center}
		\includegraphics[width=0.46\textwidth]{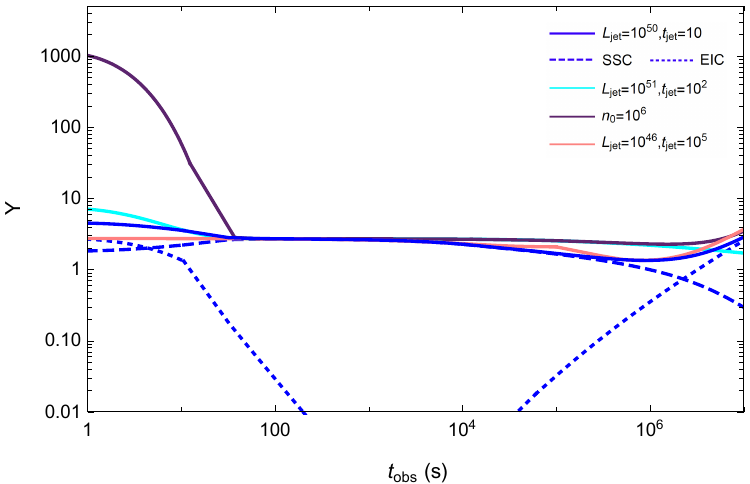}	
	\end{center}
	\caption{Comparison of the synchrotron, synchrotron
		 self-Compton scattering, and external inverse-Compton
		 scattering emission luminosity. Solid lines show the total
		 Compton–synchrotron $Y_{\text{TOT}}$, with each line
		 corresponding to various parameters in the power and 
		 duration of the jet system, as well as the ambient 
		 density. All jets are set to launch at $R_0=10^3R_g$ 
		 of an accretion disk surrounding a SMBH 
		 of mass $M=10^7 M_\odot$, and $n_0=10^8\cm^{-3}$.
		 Dashed and dotted lines denote the 
		 SSC–synchrotron $Y_{\text{SSC}}$ and EIC–synchrotron 
		 $Y_{\text{EIC}}$, respectively, for the 
		 $L_{\text{j}}=10^{50}\ergs$ and 
		 $t_{\text{j}}=10\s$ case.}	
	\label{Fig:Y}
\end{figure*}

Within the collisionless shocks generated 
by the jet-AGN gas interaction, electrons 
are compressed and accelerated to relativistic 
energy \citep[e.g.,][]{Sironi13}, of which the 
distribution can be described by a power law 
\citep{Huang03,Wang26}
\begin{equation}
\frac{d N_{e}}{d \gamma} \propto(\gamma-1)^{-p}, \quad \gamma_{m} \leq \gamma \leq \gamma_{M},
\label{eqN}
\end{equation}
where $\gamma_{m}$ and $\gamma_{M}$ are the minimum and maximum 
LF of the accelerated electrons, respectively. 
The index $p$ depends on the detailed 
microscopical processes and the magnetization of both
the jet and the ambient material, and evolves as the
shock transitions from the ultra-relativistic to
the non-relativistic regime \citep{Zhang18}. GRB
observations indicate that $p$ is event-dependent,
ranging from values below $2$ to above $3$, with a
typical value of $2.2-2.3$ consistent with 
theoretical expectations \citep{Wang15}. 
We adopt a fiducial value $p=2.3$ to describe the
electron energy distribution, while we note that $p$
may deviate from this value and evolve during jet
propagation. Assuming a fraction
$\xi_e$ of electrons are accelerated, consuming a fraction 
$\epsilon_e$ of the shock internal energy, 
the average LF of these electrons is given by
\begin{equation}
(\bar{\gamma}-1) \xi_e n_{e} m_{e} c^{2}= \epsilon_{e} \left(\Gamma-1\right) n_{p} m_{p},
\end{equation}
where $n_e$ and $n_p$ are the number density of 
electrons and protons in the shocked fluid, and 
$\Gamma$ is the LF of the post-shock gas. For 
a pairless hydrogen shock, one has $n_p/n_e=1$. 
For the distribution of Equation \eqref{eqN}, the 
average LF can be expressed as
$\bar{\gamma}-1\simeq\frac{p-1}{p-2}(\gamma_m-1)$ 
for $\gamma_m \ll \gamma_M$ and $p>2$, 
and thereby the minimum LF of electrons is
\begin{equation}
\gamma_m=\frac{p-2}{p-1}\frac{\epsilon_{e}}{\xi_{e}} \frac{m_{p}}{m_{e}}\left(\Gamma-1\right)+1,
\end{equation}
which can be extended to the Newtonian regime, where 
$\Gamma-1 \sim \frac{1}{2} \beta_\text{sh}^2$, and
\begin{equation}
\beta_\text{sh}=\left[2\frac{p-1}{p-2} (\gamma_m-1) \frac{\xi_{e}}{\epsilon_{e}}\frac{m_{e}}{m_{p}}\right]^{\frac{1}{2}}.
\label{bm}
\end{equation}
Since only relativistic electrons effectively produce 
nonthermal radiation, we adopt a critical LF
$\gamma_\text{m,cri}=2$, below which the 
electron contribution to the emission becomes negligible 
\citep{Granot06, Sironi13}. For $p=2.3$ and $\xi_e=1$,
the corresponding critical shock velocity is 
$\beta_\text{cri}=0.22 (\epsilon_{e}/0.1)^{-0.5}$.
When $\beta_\text{sh}>\beta_\text{cri}$, it can be 
considered that essentially all electrons are 
accelerated to be relativistic; conversely, 
when the shock decelerates into the deep Newtonian 
regime \citep{Huang03}, as occurs during jet propagating 
within the AGN environment, only a fraction
$\xi_e<1$ of electrons attains relativistic energies.
As shown in Equation \eqref{bm}, assuming 
$\epsilon_e$ remains approximately constant across the 
shock evolution, one has $\xi_e \propto \beta_\text{sh}^2$, 
the resulting relativistic electron fraction is given by
\begin{equation}
\xi_{e}= \text{min}[1,\left(\frac{\beta_{\text{sh}}}{\beta_\text{cri}}\right)^2],
\end{equation}
and the minimum LF of the accelerated electron 
is adjusted to $\text{min}[\gamma_m,2]$.

Besides accelerating electrons, shocks also amplify magnetic
fields. Assuming a fraction $\epsilon_B$ of the shock internal 
energy is converted into magnetic fields, the comoving-frame 
magnetic energy density is given by
\begin{equation}
U_B=\frac{B'^2}{8\pi}= 4\epsilon_{B} \left(\Gamma-1\right) \frac{\hat{\gamma} \Gamma+1}{\hat{\gamma}-1} n m_{p} c^2,
\end{equation}
where the prime denotes the quantity measured
in the the fluid comoving frame.
For the RS, retaining the expressions
for the internal energy and electron
number density of the shocked fluid, 
$\gamma_m$ and $B'$ can be calculated as 
\begin{equation}
\gamma_{m,R} = \frac{p-2}{p-1}\epsilon_{e} \frac{e_3}{n_3 m_{e} c^2}+1,
\end{equation}
and
\begin{equation}
B'_{R}=(8 \pi \epsilon_{B} e_3)^{\frac{1}{2}}.
\end{equation}
For simplicity, we adopt uniform values 
$\epsilon_{e}=0.1$ 
and $\epsilon_{B}=0.01$ across both the FS and 
RS regions, although the actual values of these 
microphysical parameters are event-dependent,
ranging over $\epsilon_{e}\in (10^{-2},0.5)$ and 
$\epsilon_{B}\in (10^{-6},0.5)$ 
\citep[e.g.,][]{Santana14}, and
may vary significantly between FS and RS 
\citep{Zhang18}.

AGN environments consist of both gas and 
photons originating from the AGN disk and corona. 
During the interaction between the jet and 
this multi-component medium, relativistic electrons 
in the shocked fluid simultaneously produce synchrotron 
radiation and inverse Compton scattering, where both 
the synchrotron photons and the AGN photons 
are upscattered. 
Thus, the nonthermal emission arises from three 
distinct radiative processes: synchrotron emission, 
synchrotron self-Compton (SSC) scattering, and external 
inverse Compton (EIC) scattering, the total luminosity 
of which is 
\begin{equation}
L_{\text {nonthermal }}= L_{\text{syn}} \left(1+Y_{\text{TOT}}\right),
\end{equation}
where $Y_{\text{TOT}}=Y_{\text{SSC}}+Y_{\text{EIC}}$ is 
the total Compton enhancement factor, with
$Y_{\text{SSC}}=L_{\text{SSC}}/L_{\text{syn}}$ and 
$Y_{\text{EIC}}=L_{\text{EIC}}/L_{\text{syn}}$ are the 
SSC and EIC luminosity ratios relative to the synchrotron 
luminosity, respectively. Derived from the same population
of electrons, the single-electron power ratio provides a 
robust proxy for the relative intensity of these three 
emissions, i.e., 
$Y_{\text{SSC}}\simeq U_\text{syn}/U_{B}=\beta_\text{sh} U_e/U_{B}(1+Y_{\text{TOT}})$
\citep{Sari01}, and $Y_{\text{EIC}}\simeq U_\text{AGN}/U_{B}$, where $ U_\text{syn}$, 
$U_e$ and $U_\text{AGN}$ are the fluid comoving-frame 
energy densities of the synchrotron radiation field, 
the relativistic electrons, and the external AGN 
radiation field, respectively \citep{Zhang18}.

Treating the AGN disk radiation as the external 
photon field, its comoving-frame energy density 
at a specific distance $R=\sqrt{R^2_0+r^2}$ from 
the central SMBH under the unidirectional light 
assumption can be calculated as 
\begin{equation}
U_\text{AGN}\simeq \Gamma^{4}(1-\beta \cos i)^{4} \cos i \frac{\eta \dot{M}c}{4\pi R^2},
\end{equation}
where $\dot{M}$ and $\eta$ are the accretion rate and 
radiative efficiency of the AGN disk, the first two terms 
correspond to the Doppler factor, $i$ is the angle 
between the disk normal and the direction of photon 
propagation from the inner disk toward the shock, 
with $\cos i =r/R$, and the third term accounts for 
the effective radiative area of the disk. 
Substituting $Y_{\text{EIC}}$, one has
\begin{equation}
Y_{\text{SSC}}=\frac{-(1+Y_{\text{EIC}})+\sqrt{(1+Y_{\text{EIC}})^2+4\beta_\text{sh}\epsilon_{e}/\epsilon_{B}}}{2}.
\end{equation}
As shown in Figure \ref{Fig:Y}, for our fiducial
$\epsilon_{B}$, the EIC component dominates the electron cooling
only during two evolutionary phases of jet propagation in the 
AGN environment: the early ultra-relativistic phase, where  
strong Doppler boosting significantly enhances $U_\text{AGN}$,
resulting in $U_\text{AGN}\gg U_{B}$ even when $\epsilon_{B}$ 
assumes an extreme value of $0.5$; and the late Newtonian phase, 
where declining shock power reduces the magnetic energy 
density, thereby suppressing synchrotron and increasing
$Y_{\text{EIC}}$. Since $Y_{\text{EIC}} \propto \epsilon_{B}^{-1}$,
lowering $\epsilon_{B}$ further amplifies the contribution of EIC
cooling. However, for most jet events, the EIC component remains
subdominant during the trans-relativistic evolutionary phase, 
unless $\epsilon_B$ is significantly reduced below fiducial value. 
In that regime, EIC becomes the dominant cooling channel, as 
synchrotron losses are suppressed due to the low magnetic field
strength. Given that AGN are extended sources,  
calculating the EIC radiation requires spatially resolved 
background emission spectra. For $\epsilon_{B}=0.01$, the 
EIC component remains subdominant across most jet evolutionary 
stages, we therefore adopt a simplified treatment by not 
modeling the EIC emission in detail, just incorporating its 
radiative contribution exclusively through $Y_{\text{TOT}}$, 
and we leave specific calculations in future work.
Meanwhile, substantial variations in either $\epsilon_{e}$
or $\epsilon_{B}$ strongly affect the electron cooling
feature when SSC dominates EIC $Y_{\text{SSC}}\gg Y_{\text{EIC}}$,
in which case
$Y_{\text{TOT}}\sim Y_{\text{SSC}} 
\sim \sqrt{\beta_\text{sh}\epsilon_{e}/\epsilon_{B}}$.

The comoving-frame peak specific synchrotron emission 
power of a single electron is \citep{Rybicki79}
\begin{equation}
P_{\nu\text {,max }}=\frac{\sqrt{3} B' e^{3}}{m_{e} c^{2}},
\end{equation} 
and its total emission power is 
\begin{equation}
P_\text{syn}(\gamma)=\frac{4}{3} \sigma_\text{T} c \gamma^2 \beta^2 U_B.
\end{equation}
Including both SSC and EIC cooling contributions, the 
radiative energy rate for an electron
with LF $\gamma$ can be written as
\begin{equation}
\frac{d \gamma}{d t} m_{e} c^{2}=-P_{\text{syn}}\left(1+Y_{\text{TOT}}\right),
\end{equation}
and the corresponding characteristic 
radiative cooling timescale is given by
\begin{equation}
t_{c}=\left|\frac{\gamma}{\dot{\gamma}}\right|=\frac{6 \pi m_{e} c^{2} \gamma}{\sigma_\text{T} c\left(\gamma^{2}-1\right) B'^{2}(1+Y_{\text{TOT}})}.
\label{tc}
\end{equation}
Equating this cooling time with the electron 
acceleration timescale 
$t_\text{acc}\simeq \gamma m_e c/e B'$, the 
maximum LF attainable by relativistic electrons 
is 
\begin{equation}
\gamma_\text{M}=\left[\frac{6\pi e}{\sigma_\text{T} B' (1+Y_{\text{TOT}})}\right]^{\frac{1}{2}},
\end{equation}
which is much larger than other characteristic LFs.
Meantime, the cooling LF,
$\gamma_c$, is defined by equating $t_{c}$ 
with the fluid comoving dynamical timescale
$t'$, i.e. \citep{Wang26}
\begin{equation}
\gamma_{c}=\frac{1}{2}\left(\bar{\gamma}_{c}+\sqrt{\bar{\gamma}_{c}^{2}+4}\right),
\end{equation}
where
\begin{equation}
\bar{\gamma}_{c}=\frac{6 \pi m_{e} c}{\sigma_\text{T} B'^{2}(1+Y_\text{TOT}) t'}.
\end{equation}
For an on-axis observer, one has $t'\simeq 2\Gamma t_{\text{obs}}$,
where $t_{\text{obs}}$ denotes the observed time measured 
from the moment of successful jet breakout.

Before escaping the emission region, synchrotron photons would
experience a frequency-dependent self-absorption by the radiating 
electrons, which modifies both the electron energy distribution
and the observed radiation spectrum. Below the characteristic absorption 
frequency $\nu'_a$, synchrotron emission is optically thick, and thus, 
photons are self-absorbed and radiation is suppressed, displaying 
a thermal-like spectral distribution in Rayleigh-Jeans regime 
\citep{Sari99}. Consequently, $\nu'_a$ can be estimated via 
equating the synchrotron flux with the blackbody flux 
\citep[e.g.,][]{Shen09, Wang26}, i.e.
\begin{equation}
I_{\nu}^{\text{syn}}\left(\nu'_{a}\right) =I_{\nu}^{\text{bb}}\left(\nu'_{a}\right) = 2 k T \frac{\nu'^{2}_{a}}{c^{2}},
\label{vaT}
\end{equation}
and
\begin{equation}
k T \sim \max[\gamma_a, \min(\gamma_c,\gamma_m) ] m_e c^2,
\end{equation}
where the LF of an electron and its characteristic
synchrotron emission frequency is connected through 
$\nu'=3 e B' \gamma^2 / 4 \pi m_e c$. In the strong 
self-absorption regime, where $\nu'_a >\nu'_c$, the 
self-absorbed photons would heat the electrons and 
prevent their cooling down to $\gamma_c$ they would 
otherwise attain. This process transforms both the 
electron energy distribution and the photon spectrum
into a quasi-thermal form below 
$\gamma_a$ on a timescale shorter than $t'$ 
\citep{Ghisellini88}, giving rise to a thermal bump in 
electron energy distribution and a corresponding 
thermal hump in emission spectrum \citep{Kobayashi04}. 
The detailed expressions of 
$\nu'_a$ are presented in Appendix \ref{va}.
The broadband synchrotron spectrum $F_{\nu}(\nu)$ and 
the electron energy distribution $N_e(\gamma)$ are 
fully determined by the relative ordering of the 
three key characteristic frequencies, 
which are presented in Appendix \ref{FN}.
For an on-axis observer, the observed photon
frequency is Doppler-boosted as $\nu \sim 2 \Gamma \nu'$.
Throughout the remainder of this paper, we adopt the
observer-frame frequency.

Considering first-order inverse Compton scattering,
given the electron energy distribution $N_e(\gamma)$ and 
the synchrotron photon spectrum $F_{\nu}(\nu_\text{s})$, 
where $\nu_\text{s}$ is the frequency of these seed photons, 
the flux spectrum of the SSC emission is given by 
\citep{Sari01, Gao13a}
\begin{equation}\label{SSC}
\begin{split}
F_{\nu,\text{SSC}}(\nu) &=\Delta \sigma_{\text{T}} \int_{2}^{\infty} d \gamma \frac{d N_e}{d \gamma} \int_{0}^{x_{0}} d x F_{\nu}(x) \\
&=\sigma_{\text{T}} \int_{2}^{\infty} d \gamma \frac{d \Sigma_e}{d \gamma} \int_{0}^{x_{0}} d x F_{\nu}(x),
\end{split}
\end{equation}
where $\Delta$ is the thickness of the emission 
region, $\Sigma$ is the column density of the 
electrons, $x=\nu/4\gamma^2\nu_\text{s}$ and 
$x_0=0.5$. In the strong absorption regime, 
inverse Compton scattering involving seed 
photons with $\nu_\mathrm{s} < \nu_a$ and 
electrons with LF $\gamma < \gamma_a$ is 
suppressed, as both components are thermalized 
and no longer participate in nonthermal 
scattering processes.

\section{nonthermal Emission Properties}
\label{Sec3}
\begin{figure*}
	\begin{center}
		\includegraphics[width=0.49\textwidth]{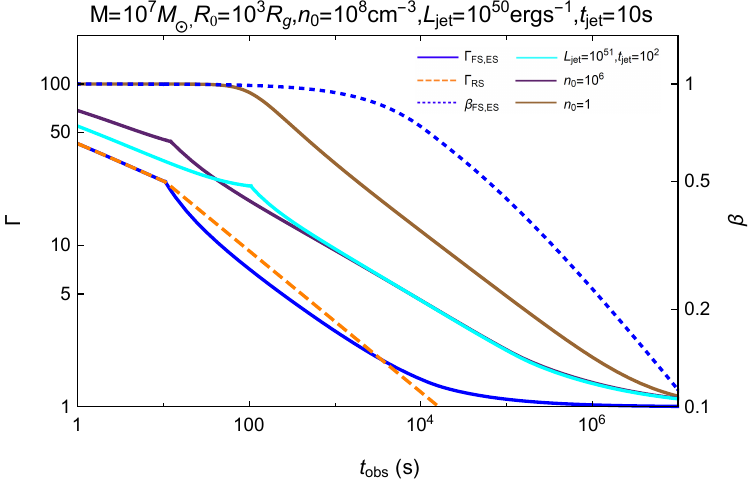}
		\includegraphics[width=0.49\textwidth]{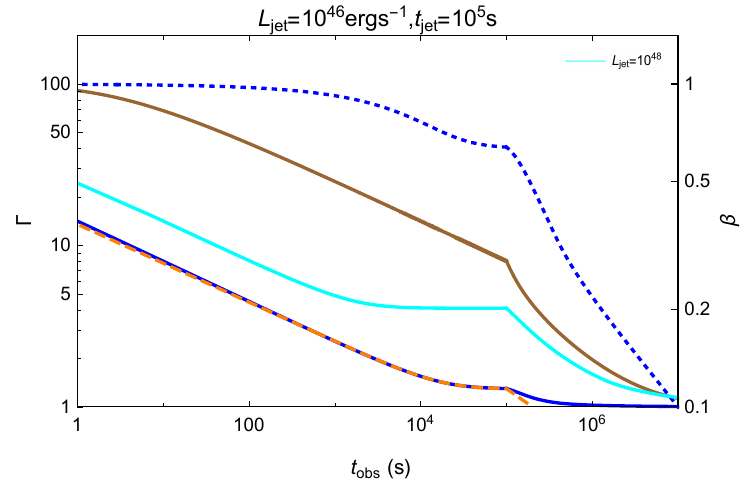}
	\end{center}
	\caption{Dynamics of jets propagating within the AGN environment. 
	The left panel shows the dynamical evolution of a jet with $L_{\text{j}}=10^{50}\ergs$ and $t_{\text{j}}=10\s$, where the 
	blue line, orange dashed line, and blue dotted line represents 
	the LF of forward shock, LF of reverse shock, and the velocity 
	of forward shock, respectively. The cyan and purple line exhibits 
	the LF of forward shock, changing the system parameter to a more 
	powerful and long-lived jet with $L_{\text{j}}=10^{51}\ergs$ and 
    $t_{\text{j}}=100\s$, and a lower-density ambient medium 
    $n_0=10^6\cm^{-3}$, respectively. For comparison with the canonical 
    environment, the interstellar medium case is also presented, 
    where $n_0=1\cm^{-3}$. The right panel shows a long-lived jet case 
    with $L_{\text{j}}=10^{46}\ergs$ and $t_{\text{j}}=10^5\s$, where 
    the cyan line changes the jet power to $L_{\text{j}}=10^{48}\ergs$. }
	\label{Fig:dyn}
\end{figure*}

\begin{figure*}
	\begin{center}
		\includegraphics[width=0.46\textwidth]{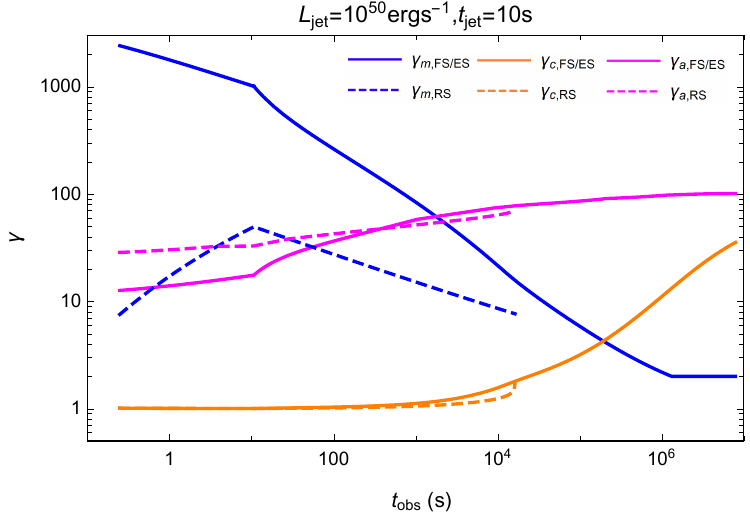}
		\includegraphics[width=0.46\textwidth]{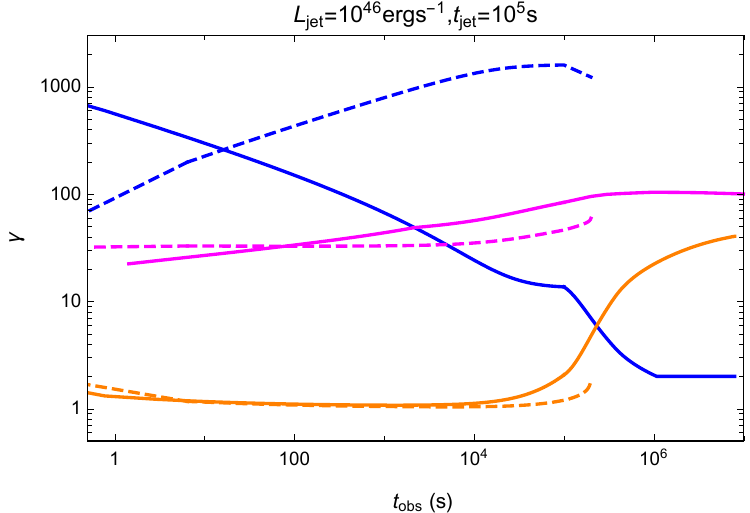}
		\includegraphics[width=0.46\textwidth]{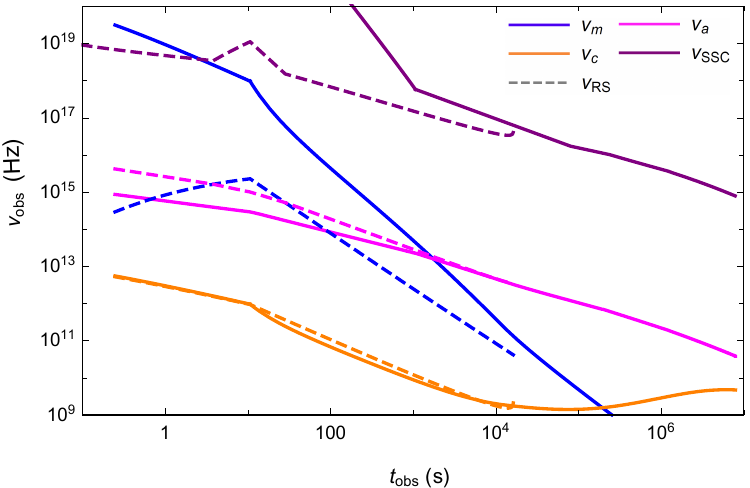}
		\includegraphics[width=0.46\textwidth]{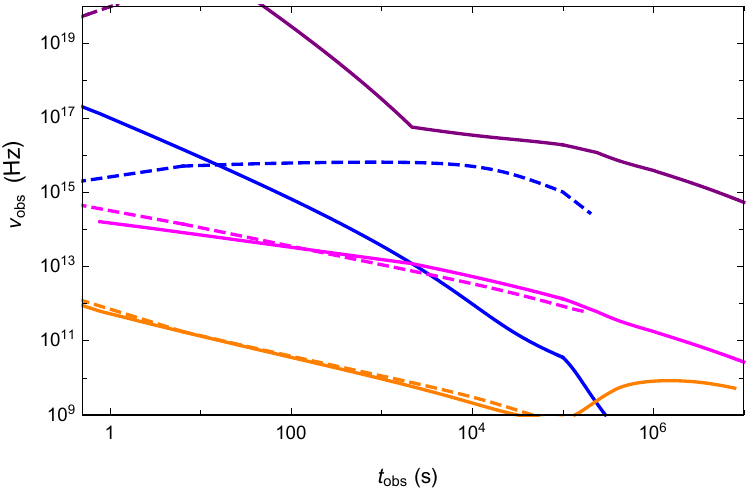}
	\end{center}
	\caption{Characteristic Lorentz factors and emission 
		frequencies of relativistic electrons. In each 
		panel, the solid and dashed lines represent 
		the parameters in FS and RS. The blue,
	    orange, and magenta line denotes the minimum
        LF $\gamma_m$, the cooling LF $\gamma_c$, and 
        the synchrotron self-absorption LF $\gamma_a$, 
        respectively. In addition to the associated 
        synchrotron frequencies, the critical frequencies 
        for SSC scattering are shown in the 
        $t_{\rm obs}$–$\nu_{\rm obs}$ panels.
        The RS-related parameter evolution is 
        artificially truncated at $\gamma_3 = 1$.  
        The left panels show a $L_{\text{j}}=10^{50}\ergs$ 
        and $t_{\text{j}}=10\s$ system, while the right
        panels show a $L_{\text{j}}=10^{46}\ergs$ and 
        $t_{\text{j}}=10^5\s$ system.}	
	\label{Fig:gamma}
\end{figure*}

By combining the dynamical and radiative 
equations presented in Sections \ref{dyn} 
and \ref{rad}, we model the co-evolution 
of jet propagation dynamics and nonthermal 
radiation in an AGN ambient medium. 
For simplicity, we adopt 
$\epsilon \sim \epsilon_e$, and examine 
three representative jet systems that 
escape the AGN disk, which collectively 
span the key parameter space governing jet–medium 
interaction—namely, a powerful uncollimated 
jet, a long-lived free jet, 
and an under-accelerated jet 
at breakout—to elucidate 
their distinct dynamical and radiative 
properties.

\subsection{Powerful uncollimated jet}
\label{SJ}
\begin{figure*}
	\begin{center}
		\includegraphics[width=0.47\textwidth]{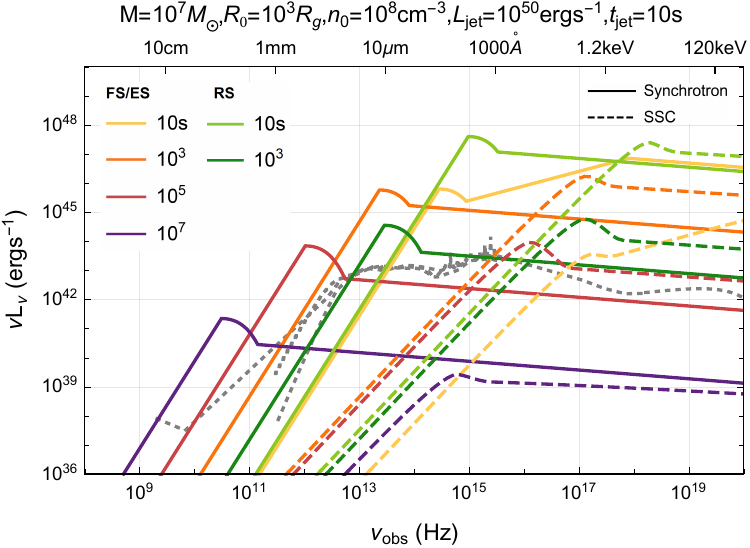}
		\quad
		\includegraphics[width=0.47\textwidth]{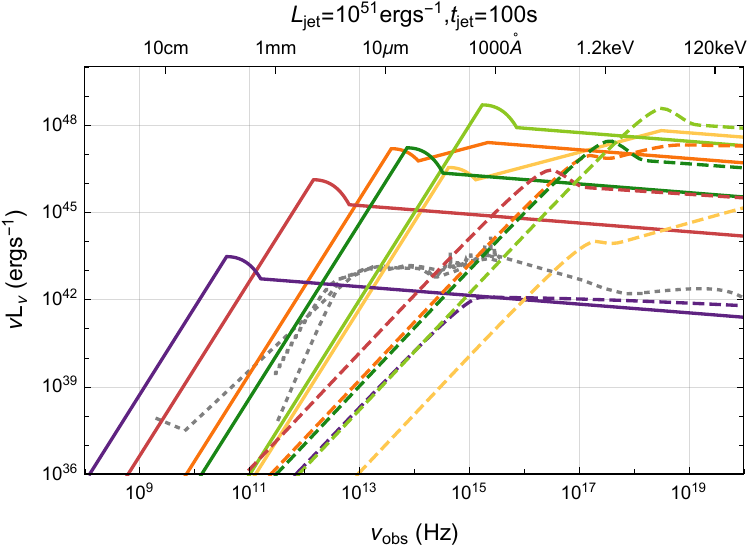}
		\includegraphics[width=0.47\textwidth]{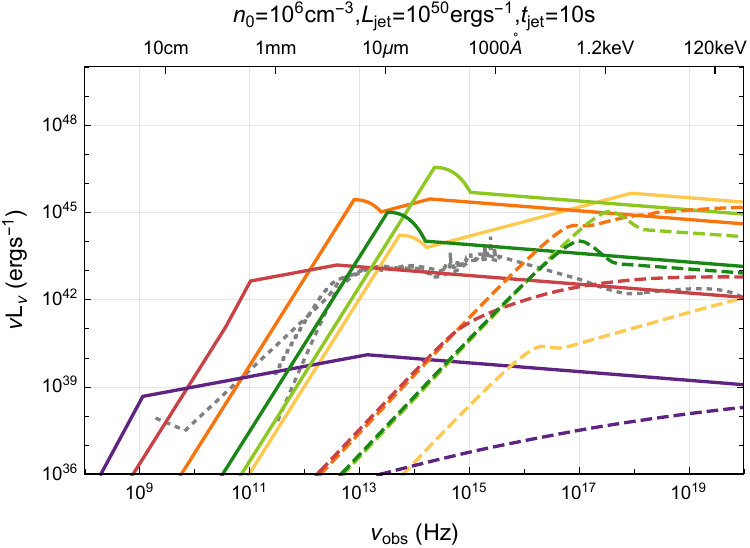}
	\end{center}
	\caption{Spectral energy distributions of nonthermal 
		emissions for GRB jet cases. Each panel corresponds 
		to different parameter: $L_{\text{j}}=10^{50}\ergs$ 
		and $t_{\text{j}}=10\s$, $L_{\text{j}}=10^{51}\ergs$ 
		and $t_{\text{j}}=100\s$, and $n_0=10^6\cm^{-3}$. 
		Emissions from both the forward and reverse shock 
		at various observer times are presented, where 
		solid lines denote synchrotron radiation, while 
		the dashed lines denote SSC emission. As a benchmark 
		for comparison, the mean spectral energy distribution 
		of radio-quiet quasars from \cite{Shang11} is shown 
		as the grey dotted line. The far-infrared band 
		constitutes a key radiation channel for jet systems;
		however, the composite SED from \cite{Shang11}
		lacks far-infrared data. Two infrared-specific 
		spectral SED with the different extent of 
		subtraction for the galaxy contribution are 
		adopted, where the middle dotted line 
		corresponds to the SED from \cite{Symeonidis16}, 
		and the bottom dotted line corresponds to the SED 
		from \cite{Lyu17}.
    }   
	\label{Fig:SED-GRB}
\end{figure*}

A fully uncollimated jet at breakout 
from the AGN disk requires a power 
$L_{\text{j}}\gtrsim10^{48}\ergs$, as shown
in Figure \ref{Fig:col}. GRB jets with 
$L_{\text{j}}\geqslant 10^{50}\ergs$ 
provide a compelling astrophysical 
realization of such disk-embedded relativistic 
jets. As shown in Figure 6 of \cite{Chen25},
at the inner region of the AGN disk
within $R_0\lesssim O(10^{3})R_g$, long 
GRB jets with $t_{\text{j},0}\sim 10-100 \s$ 
can successfully break out of the disk.
Though short GRB jets with $t_{\text{j},0}<2\s$ 
require breakout from a more inner disk 
region—typically associated with lower-mass 
SMBHs where the AGN disk scale height is 
reduced, a plausible hyper-Eddington 
accretion process would open a low-density 
circum-binary cavity 
penetrating through the disk prior to 
binary merger \citep{Chen23}, thereby 
preventing jet choking and enabling its
sustained relativistic propagation
\citep{Zhang24, Yuan25}. Here, we construct 
a representative short-duration 
uncollimated jet event to investigate 
its dynamical and radiative properties: 
a relativistic jet with power 
$L_{\text{j}}=10^{50}\erg\s^{-1}$ 
launches at $R_0=10^3R_g$ of an 
AGN disk surrounding a SMBH 
of mass $M=10^7 M_\odot$
\footnote{In the AGN merger channel, 
a substantial fraction of binary compact
object (e.g., BBH) mergers are expected to occur 
within the AGN disks surrounding SMBHs of
mass $M=O(10^7) M_\odot$ \citep{Rowan25}.}, 
successfully escapes the disk and injects 
into the AGN ambient medium, with an initial 
LF $\Gamma_\text{j}=100$ and a reduced
duration $t_\text{j}=10 \s$.

While the density of the AGN wind 
medium is lower than that of the underlying
AGN disk, it remains substantially higher 
than the density of typical interstellar 
or intergalactic environments. Consequently,
the relativistic jet experiences strong 
and rapid ram-pressure deceleration 
during propagation. 
As shown in the left panel of Figure 
\ref{Fig:dyn}, the bulk LF of the shock 
has begun to decline significantly during 
the early phase of coupled forward–reverse 
shock evolution. Following RS crossing at 
$t_\text{cross} \sim t_{\text{j}}$,
deprived of further jet energy injection, 
the FS undergoes markedly accelerated 
deceleration relative to the pre-crossing 
phase. As a result, the external shock 
evolves to a non-relativistic velocity 
on a timescale of $O(10^4)\s$. 
In contrast, when the same jet propagates 
through a uniform ISM 
with number density $n_0=1\cm^{-3}$, 
the forward shock maintains a significantly 
higher bulk Lorentz factor and remains 
relativistic at the same $t_{\text{obs}}$.
Increasing the jet power and duration
—e.g., to $L_{\text{j}}=10^{51}\ergs$ and 
$t_{\text{j}}=100\s$—or decreasing the 
AGN ambient density—e.g., to
$n_0=10^6 \cm^{-3}$—mitigates FS deceleration: 
in the former case, ram-pressure driving is 
stronger and more sustained; in the latter, 
ambient inertia provides weaker braking.

In the strong shock driven by jet-AGN medium
interaction, the evolution of nonthermal 
radiation exhibits distinguishing features, 
as shown in Figure \ref{Fig:gamma}. 
First, synchrotron emission remains 
persistently in the strong self-absorbed 
regime where $\gamma_a>\gamma_c$, owing to the 
extremely high ambient number density
\footnote{Self-absorption is more extreme
in the AGN disk environment \citep{Wang22}.}. 
At early times, 
$\gamma_c \sim 1$ with $\bar{\gamma}_c \ll 1$, 
indicating that radiative cooling is extremely
efficient, such that only a small fraction of 
the relativistic electron population contributes 
to the instantaneous emission \citep{Rahaman25}.  
Second, given that 
$\gamma_m -1 \propto \Gamma -1$,
the rapid deceleration of the FS
drives a corresponding decline in $\gamma_m$.
When $t_{\text{obs}}>10^3\s$, $\gamma_m$
drops below $\gamma_a$, shifting the 
dominant emission mechanism to 
synchrotron self-absorption thermalization. 
At late times, the Newtonian shock 
accelerates electrons inefficiently, 
resulting in $\gamma_m=2$ and a substantial 
reduction in the fraction of radiating 
relativistic electrons. Third, the short 
propagation distance over which deceleration 
occurs yields a high density of the shocked 
jet material, which consequently raises 
$\gamma_a>\gamma_m>\gamma_c$ in the RS.

The evolution of the three characteristic
synchrotron frequencies—$\nu_m$, $\nu_a$, 
and $\nu_c$—which jointly govern the 
shape of the broken power-law spectrum, 
is shown in Figure \ref{Fig:gamma}. Also 
displayed is the critical frequency of the 
SSC emission, defined as
$\nu_{\text{SSC}}=4x_0\max[\gamma_a,\gamma_m,\gamma_c]^2
\max[\nu_a,\nu_m,\nu_c]$ \citep{Sari01, Gao13a}. 
All these frequencies decrease with time, 
systematically shifting the spectral energy 
distribution (SED) toward lower energies. 
The SED exhibits distinct features across 
epochs, as shown in Figure \ref{Fig:SED-GRB}.
Overall, as $\gamma_a>\gamma_c$, 
self-absorption process heats the cooling 
electrons up and produces a thermal
hump around $\nu_a$ \citep{Kobayashi04}.
At early time $t_{\text{obs}}=10\s$, the
FS SED displays two spectral breaks at $\nu_a$ 
and $\nu_m$, peaking at $\nu_m$. 
In contrast, RS generates a more luminous 
SED, peaking at $\nu_a$. SSC emission from
the RS peaks at a lower frequency than that
from the FS, yielding low-frequency RS-SSC
flux comparable to the FS synchrotron 
component. During the intermediate time
$t_{\text{obs}}=10^3\s$ and $10^5\s$, the 
$\nu_m$ break gradually approaches $\nu_a$ 
and ultimately vanishes. Concurrently, the
FS synchrotron luminosity  
declines continuously, accompanied by a 
downward shift of the peak frequency $\nu_a$. 
As both $\nu_a$ and the corresponding 
LF $\gamma_a$ decrease, the FS SSC 
spectrum softens and its luminosity diminishes.  
At this epoch, synchrotron and SSC emission 
from the RS are both subdominant relative to 
their FS counterparts.
At late time $t_{\text{obs}}=10^7\s$, the 
radiation evolution follows the same 
trend: the total flux continues to fade,
and the spectral peak shifts progressively 
toward lower frequencies.

As the jet resides in AGN, its 
emitted radiation must be assessed 
relative to the background emission
(gray-dotted line in Figure 
\ref{Fig:SED-GRB}). In our fiducial
case, the jet emission exceeds the 
AGN background across multiple wavelength
bands at early phases; at late times, 
although it drops below the background 
at high frequencies, the jet remains 
the dominant contributor to the 
transient radio-IR emission from
its host AGN. Increasing the jet 
power and duration can produce a 
longer-lasting, multiband-observable 
emission (second panel in Figure 
\ref{Fig:SED-GRB}), its SED evolution 
closely resembles that of 
the fiducial case. A low-density AGN 
environment markedly alters 
the jet emission properties 
(third panel in Figure \ref{Fig:SED-GRB}). 
At early times,
e.g., $t_{\text{obs}}=10\s$, the
lower magnetic energy density 
downstream of the FS and the higher
bulk LF of the shock jointly yield
a large $Y_{\text{EIC}}$. Consequently, 
EIC process strongly suppresses 
both synchrotron and SSC 
emissions, leading to significantly 
reduced broadband luminosity. At late 
times, $\nu_a$ decreases significantly
in a tenuous medium, shifting the 
broken power-law spectrum to 
follow $\nu_m<\nu_a<\nu_c$ regime,
markedly distinct from that observed 
in high-density environments. 
Therefore, nonthermal emission from 
the jet system serves as a powerful 
probe of the gas profile in the AGN 
environment.  

\subsection{Long-lived free jet}
\label{LJ}
\begin{figure*}
	\begin{center}
		\includegraphics[width=0.47\textwidth]{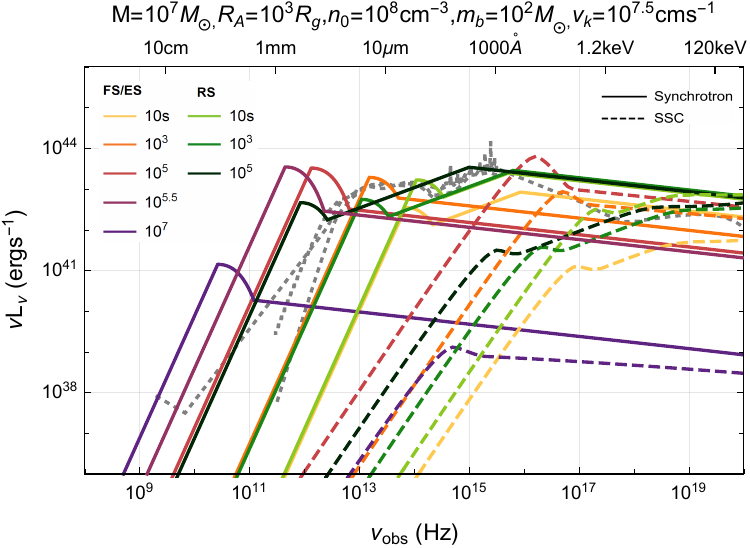}
		\quad
		\includegraphics[width=0.47\textwidth]{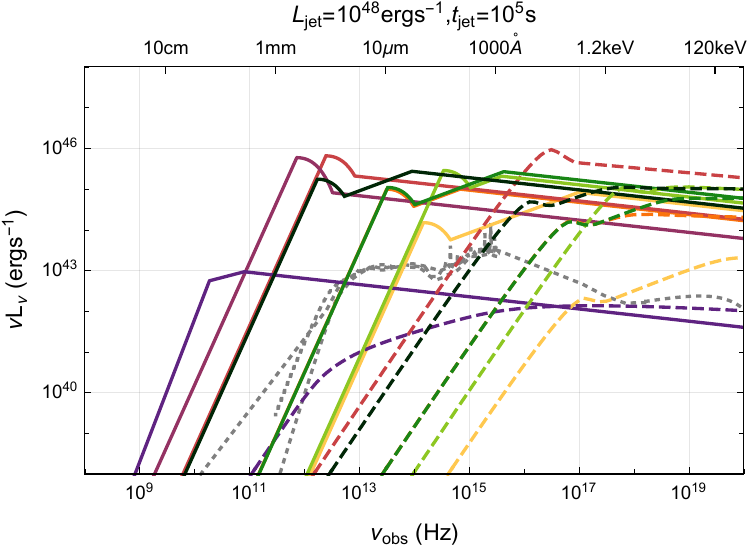}
	\end{center}
	\caption{Spectral energy distributions of nonthermal emissions 
		for BBH remnant-driven jet cases. The left panel depicts a 
		relativistic jet with $L_{\text{j}}=10^{46}\ergs$ 
		and $t_{\text{j}}=10^5\s$, consistent with the jet launched
		in a BBH merger system of mass $m=100M_\odot$, where the 
	    remnant BH receives a kick velocity 
	    of $v_{\text{k}}=10^{7.5}\cm \s^{-1}$. The right panel 
        corresponds to a more powerful jet with 
        $L_{\text{j}}=10^{48}\ergs$.}	
	\label{Fig:SED-BH}
\end{figure*}

The duration of a long-lived free
jet satisfies $t_{\text{j},0}\gg t_{\text{bre}}$.
Jets launched by BBH mergers embedded in 
AGN disk fulfil this condition 
and therefore constitute a representative 
example of this jet class. Due to the 
anisotropy of gravitational-wave emission, the 
remnant black hole acquires a recoil velocity 
and subsequently moves through the AGN disk, 
where it accretes ambient magnetized gas and 
launches a jet \citep{Chen24}. The jet 
luminosity can be estimated as 
\citep{Kaaz23,Kim25}
\begin{equation}
\begin{split}
L_{\text{j}}&=\eta \dot{M}_{\text{BHL}} c^2 \\
&=1.3\times 10^{46}\ergs\left(\frac{\eta}{0.05}\right)
\left(\frac{\dot{M}_{\text{BHL}}}{2.3\times 10^{26}\text{~g}\cm^{-3}}\right),
\end{split}
\end{equation}
where $\eta$ denotes the effective 
energy conversion efficiency, 
$\dot{M}_{\text{BHL}}$ is evaluated 
for an accretion system that a BH 
of mass $m=100 M_\odot$ moves at 
$v_{\text{k}}=10^{7.5}\cm\s^{-1}$ 
through the AGN disk at $R_0=10^3R_g$ 
around a SMBH of mass $M=10^7M_\odot$ 
\citep{Chen24}. 
The jet duration is highly sensitive to the 
magnetic properties of the AGN disk. 
Driven by the magnetorotational instability, 
the disk magnetic field is turbulent, with 
field directions randomized but exhibiting
approximate coherence over $H_\text{d}$ scale 
\citep[e.g.,][]{Salvesen16}. For a given
magnetic field inclination angle related to the BH
spin, the duration of an individual jet corresponds
to the accretion timescale $t_{\text{BHL}}$ 
\citep{Kim25}. Besides, the jet is quenched 
once the ambient magnetic field reorients or the 
remnant BH escapes the AGN disk; 
both processes occur on a characteristic timescale
$t_{\text{k}}\sim H_\text{d}/v_{\text{k}}$. 
Consequently, the jet duration is 
$t_{\text{j},0}=\min[t_{\text{BHL}}, t_{\text{k}}]$.
As shown in Figure 2 of \cite{Chen24}, for launching 
radii $R_0 \gtrsim \text{a few hundred } R_g$, we 
have $t_{\text{BHL}} < t_{\text{k}}$. 
Thus, we conservatively adopt  
\begin{equation}
\begin{split}
t_{\text{j},0}&\simeq t_{\text{BHL}}=\frac{G m}{v_{\text{k}}^3} \\
&=4.1\times 10^5\s \left(\frac{m}{100 M_\odot}\right) \left(\frac{v_{\text{k}}}{10^{7.5}\cm\s^{-1}}\right)^{-3}.
\end{split}
\end{equation}
As shown in Figure 4 
of \cite{Chen25}, the jet breakout time 
is typically less than one day. 
Therefore, a kicked remnant—acting
as the central engine—remains active
over an extended period during its driven
jet propagating through the AGN environment,
continuously injecting unshocked 
material to push the FS and RS.  
Here, we investigate this representative 
jet, adopting $L_{\text{j}}= 10^{46}\ergs$,
$t_{\text{j}}= 10^{5}\s$, and 
$\Gamma_{\text{j}}=100$.

The FS-RS evolution phase persists for
$t_{\text{cross}} \sim t_{\text{j}}$. 
When propagating through a low-density
ISM, the FS can maintain 
relativistic velocity over long 
timescales, as shown in the right 
panel of Figure \ref{Fig:dyn}. 
In contrast, within the AGN environment,
owing to high ambient gas density and 
moderate jet power, the LF of the shocked 
fluid decreases significantly over time: 
it falls to $\sim 10$ at early time 
$t_{\text{obs}}=1\s$, and even becomes 
non-relativistic despite continued  
energy injection from the ongoing jet. 
Once all the jet materials are shocked, 
the FS undergoes an extreme braking, as 
evidenced by a sharp decline in 
velocity $\beta$. Another distinctive 
feature is that the LF of the shocked 
fluid asymptotically flatten once the 
FS–RS system propagates over sufficiently 
large distances, where the AGN
environment exhibits wind density 
profile. This behavior is prominent in jet 
system with $L_{\text{j}}= 10^{48}\ergs$,
arising from the identical radial 
dependence—$n(n_{\text{j}}) 
\propto r^{-2}$—of the AGN and jet density, 
leading to an approximately constant 
LF, as demonstrated by Equation 
\eqref{g2}. The strong shock
efficiently converts most of the jet 
kinetic energy into the internal 
energy of the shocked fluid, which 
subsequently powers substantial
nonthermal emission.  

As shown in Figure \ref{Fig:gamma}, 
the overall evolution of the characteristic 
LFs and frequencies of synchrotron emission 
closely mirrors that of the 
powerful uncollimated jet system analyzed 
in the previous section. Moreover, 
for the FS, its LF $\gamma_m$ displays 
a flattened temporal evolution during the 
late FS-RS stage, attributable 
to the near-constant shock velocity 
sustained in the AGN wind medium. For the RS,
$\gamma_m$ is larger than the self-absorption 
LF $\gamma_a$, on account of its LF 
$\gamma_{34}$ attaining a high value,
which then produces a large post-shock 
electron energy density $e_3$.

For the long-lived jet case, the most 
prominent feature is the prolonged persistence 
of the RS. As during the FS-RS evolution stage, 
the peak synchrotron flux of RS fluid exceeds
that of FS \citep{Kobayashi03}; consequently, 
the radiation is dominated by the RS over a 
broad spectral range, shown in Figure
\ref{Fig:SED-BH}. The SED at various epochs 
generally exhibits a thermal hump at $\nu_a$, 
arising from the strong self-absorption.   
For the $L_{\text{j}}= 10^{46}\ergs$ jet, 
the emission remains comparable to or 
exceeds the background at high frequencies 
during the FS-RS evolution phase, which is
predominantly powered by the RS, with
luminosity in these bands remaining 
approximately constant, arising from a 
combination of three effects: only a 
fraction of electrons radiate instantaneously,
progressive strengthening of the RS, and 
Doppler boosting.
At later times, as $\nu_a$ evolves to lower 
frequencies, where the background radiation 
is comparatively weaker, the emission becomes
detectable. Increasing the jet power, 
e.g., to $L_{\text{j}}= 10^{48}\ergs$, elevates 
the overall luminosity of the jet SED, rendering 
it brighter than the background radiation 
across multiple bands over an extended duration.  
Moreover, during the FS-RS evolution phase, 
the more powerful jet sustains higher shock 
velocity, leading to a slower decay of the FS
emission and a late modest re-brightening,
attributable to the near-constant shock 
velocity combined with an expanding shock 
surface area. After the RS crossing, comparison 
of the SEDs at $t_{\text{obs}}=10^5\s$, 
$10^{5.5}\s$, and $10^{7}\s$ for both cases 
reveals that, the emission 
undergoes a slightly re-brightening episode, 
driven by the enhanced deceleration of FS 
following the cessation of jet energy injection, 
which thereby facilitates a more efficient 
conversion of ejecta kinetic energy into radiation. 

\subsection{Under-accelerated jet at breakout}
Unlike the two jet classes analyzed above, 
when the initial jet duration satisfies 
$t_{\text{j,cri}}<
t_{\text{j},0}\lesssim t_{\text{bre}}$,
at breakout, the emerging jet is 
initially shocked and photon-dominated 
with most of its energy residing 
in the radiation field, which subsequently 
accelerates the jet material.
As shown in Figure \ref{Fig:col}, 
for an initial jet LF of $\Gamma_0 = 100$,
the critical baryonic loading 
satisfies $\eta_b<20$, meaning that
the re-accelerating jet fails to 
reach its initial LF, 
and photons escape before the 
acceleration completes, with 
$1-\eta_b/\Gamma_0 >0.8$ 
of the jet initial energy consumed 
in the form of radiation, producing
a transient flare. The duration of this 
emission is comparable to that of 
the energy injection into the shocked 
jet, which can be estimated as
\citep{Nakar17}
\begin{equation}
t_{\text{j,sh}} \approx t_{\text{j},0},
\end{equation}
and its isotropic equivalent 
luminosity is
\begin{equation}
L_{\text{j,sh}} \approx \frac{2 L_{\text{j}}}{\theta_0^2}.
\end{equation}
The radiation temperature
can be taken as 
\citep{Nakar12}
\begin{equation}
T_{\text{j,sh}} \sim 100\, \text{keV},
\end{equation}
regulated by the positron-electron 
pair production.

As only a fraction of the jet total energy
is deposited as kinetic energy,
both the initial power and LF of the 
emerging jet are reduced compared to those of 
a free jet propagating through 
the AGN environment. This partially 
accelerated jet can nonetheless produce 
nonthermal radiation via shock 
interaction process with the ambient
AGN medium, consistent with the scenarios 
discussed in Section \ref{SJ} and 
\ref{LJ}. However, the lower jet power 
results in weaker radiative luminosity. 
Meanwhile, the jet decelerates more 
rapidly, leading to a faster-evolving 
SED compared to a free jet of 
identical duration and full acceleration. 
Moreover, the lower LF implies a smaller 
Doppler factor, which
further reduces the early observed 
luminosity. As a result, the post-flare 
emission from the partially accelerated jet
is observationally fainter and thus less
detectable than that from its fully 
accelerated counterpart.

\section{Detectability}
\label{Sec4}
For an AGN, the successful detection 
of emission from a stellar-object-driven jet 
depends not only on progressive 
improvements of telescope sensitivity, 
but also critically on whether the 
source intrinsic luminosity exceeds 
that of the background radiation. 
To assess detectability, we investigate 
multi-wavelength light curves of the 
targeted jet systems, taking the 
corresponding AGN background radiation 
as the baseline.

\subsection{Gamma-ray burst}
\begin{figure*}
	\begin{center}
		\includegraphics[width=0.47\textwidth]{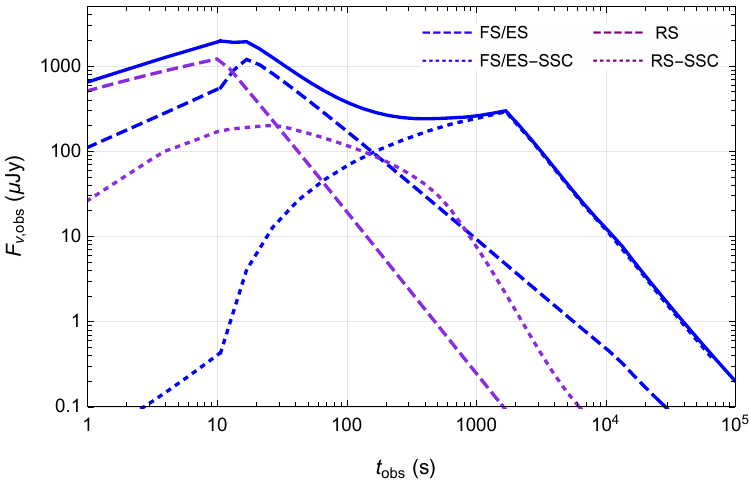}
		\includegraphics[width=0.47\textwidth]{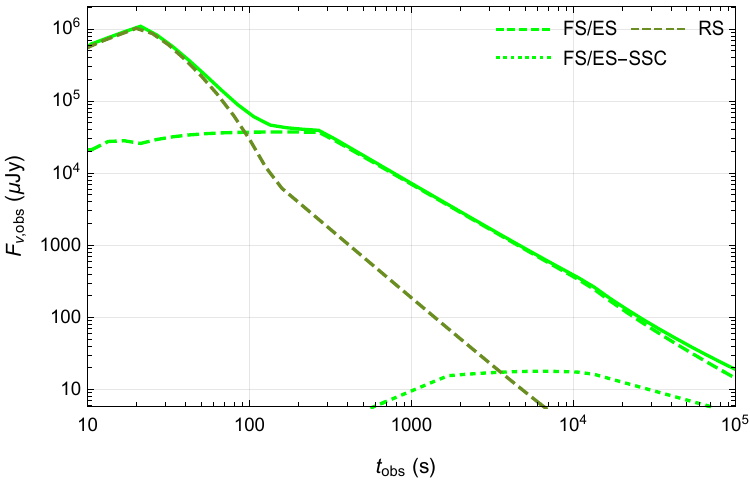}
		\includegraphics[width=0.47\textwidth]{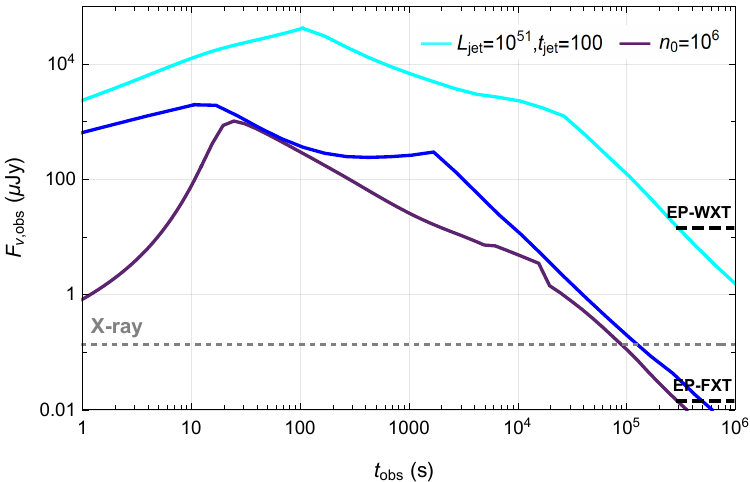}
		\includegraphics[width=0.47\textwidth]{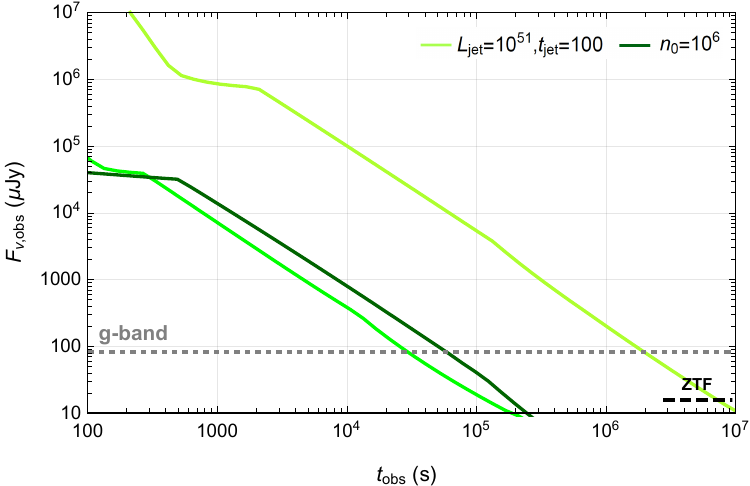}
	\end{center}
	\caption{X-ray (left panels) and optical (g-band, right panels) 
	light curves of GRB jet emission.
	The jet is set as $L_{\text{j}}=10^{50}\ergs$ and $t_{\text{j}}=10\s$.
    The upper panels shows the total observed flux evolution, decomposed 
    into contributions from the FS and RS synchrotron components, and the 
    corresponding SSC components. The lower panels display the light curves 
    for varying parameters,
    where cyan and light green lines represent the 
    $L_{\text{j}}=10^{51}\ergs$ and $t_{\text{j}}=100\s$ case, purple 
    and dark green lines correspond to the $n_0=10^6\cm^{-3}$ case. The
    grey dashed lines indicate the flux level of AGN background in the 
    respective band. Also shown are the detection sensitivities of the 
    soft X-ray telescope Einstein Probe (EP) for $10\text{~ks}$ 
    of exposure time \citep{Yuan22}, and the optical telescope Zwicky 
    Transient Facility (ZTF) for $30\s$ of exposure time \citep{Bellm14}.}	
	\label{Fig:Flux-GRB}
\end{figure*}

\begin{figure*}
	\begin{center}
		\includegraphics[width=0.47\textwidth]{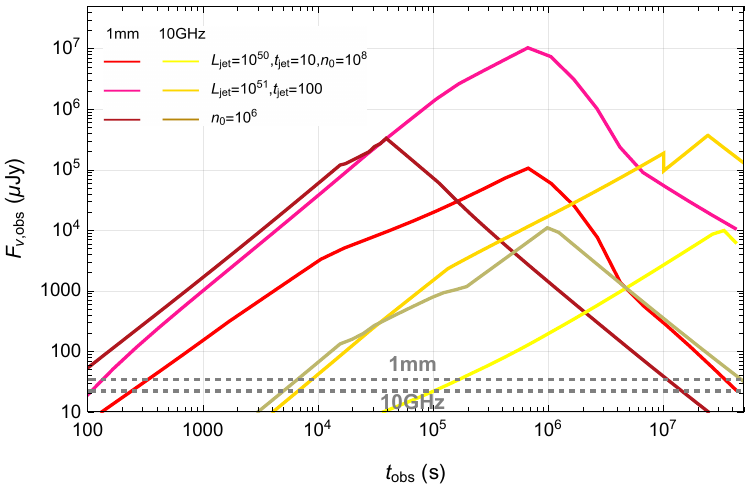}
	\end{center}
	\caption{Same as Figure \ref{Fig:Flux-GRB}, but for
		infrared ($1\text{~mm}$) and radio ($10\text{~GHz}$) 
		light curves of GRB jet emission.}	
	\label{Fig:Flux-LF}
\end{figure*}

As revealed by the SED as shown in 
Figure \ref{Fig:SED-GRB}, the GRB jet 
emission outshines the background 
across multiple observational bands. 
Moreover, its total luminosity arises 
from the superposition of multiple 
radiation components—synchrotron emission
and SSC scattering from both FS and 
RS—rendering complex structure in the light 
curves across different bands. 

In the soft X-ray band (e.g., $1 \text{~keV}$), 
the dominant radiation mechanism responsible 
for the detected emission evolves across 
observational epochs, as shown in the upper left panel of 
Figure \ref{Fig:Flux-GRB}. During the early 
FS–RS evolution phase, synchrotron emission from 
the RS fluid dominates the observed flux. 
After RS crossing, its emission decays rapidly, 
and FS synchrotron component becomes dominant.
Subsequently, as the critical frequency 
$\nu_{\text{SSC}}$ shifts through 
$1 \text{~keV}$, FS-SSC emission surpasses 
the fading synchrotron component, becoming the 
primary contributor to the soft X-ray flux. 
In contrast, RS-SSC emission remains the least 
luminous among the four components. Accordingly,
in addition to the characteristic rise and decay 
phases, the light curve exhibits an approximately 
plateau-like structure during the transition from the 
synchrotron-dominated to the SSC-dominated regime
(a similar feature is reported in \cite{Zhao26}).
As shown in the lower left panel of Figure \ref{Fig:Flux-GRB},
the powerful GRB jet drives soft X-ray emission 
that rises significantly above the AGN background, 
yielding a detectable flare persisting for approximately 
one day. Increasing the jet power or duration
enhances the peak luminosity of soft X-ray flare, 
and extends its observable time window to 
several days. Decreasing the ambient density 
significantly suppresses the early-time luminosity 
of both synchrotron and SSC emission. The AGN 
photons consume nearly all the energy of relativistic 
electrons via EIC scattering in both FS and RS, 
causing an early flux dip to emerge in 
the light curve\footnote{EIC scattering would produce 
intense soft X-ray emission.}. Meanwhile, 
FS-SSC emission becomes dominant at a later 
epoch, preventing the emergence of a discernible 
plateau in the light curve.  

As shown in the right panels of 
Figure \ref{Fig:Flux-GRB},
in the optical band (e.g., g-band), 
synchrotron emission from both the 
FS and RS jointly accounts for the 
gradual decay of the early-time light 
curve. Subsequently, the rapid fading of 
the RS component induces a steeper decay 
phase. Moreover, during the very early 
stage, strong self-absorption within the RS 
produces an optical thermal hump. SSC 
emission from both the FS and RS is too 
faint to yield a detectable contribution 
to the optical light curve.  
Given that this band is the dominant band 
of AGN background emission—due to its high 
intrinsic luminosity—the jet system can 
only produce flare with duration shorter 
than one day. A more powerful jet or
a longer jet active time can concurrently
enhance both the peak luminosity and 
duration of the optical flare.

In the infrared and radio band
(e.g., $1 \text{~mm}$ and $10 \text{~GHz}$),
radiation is dominated almost exclusively 
by synchrotron emission from the FS.
As the peak frequency of SED shifts 
into the infrared and radio bands 
during late-time epochs, the corresponding 
flares peak at $\sim 10^6\s$ and 
$\sim 10^7\s$, respectively, as shown in
Figure \ref{Fig:Flux-LF}. The prominent 
decay structure observed in the light 
curves arises because $\nu_a$ and its 
driven thermal hump cross the observed 
frequency band, causing a plunge in 
emission luminosity. In a low-density 
environment (e.g., $n_0=10^6 \cm^{-3}$), 
the FS emission transitions from the 
strong-absorption regime to the 
weak-absorption regime, where 
$\nu_a<\nu_c$, at $\sim 10^4\s$. 
Subsequently, the thermal hump disappears, 
and the SED peak frequency downshifts into 
$1 \text{~mm}$ and $10 \text{~GHz}$ at an 
earlier time. Concurrently, the peak flux 
densities in both bands are comparable to
those in the high-density case.
The flares produced by the jet system 
in these bands significantly outshine 
the AGN background, making them highly 
detectable transient signals. If the 
AGN disk-embedded BNS/NS-BH merger GW 
signal is detected, as the time delay of 
these EM transients is shorter than 
$O(10^6)\s$, the association between 
the GW-EM signals can be robustly confirmed.
Moreover, this radiation component can 
induce transient modifications to the 
AGN SED on timescale of months to years 
—for instance, temporarily enhancing 
the radio-band luminosity sufficiently 
to reclassify a radio-quiet AGN as 
radio-loud.

Overall, GRB jets propagating through 
the AGN environment act as multi-wavelength 
emitters, producing detectable radiations 
across a broad spectral range, which exceed
the sensitivity threshold of diverse 
telescopes.

\subsection{BBH merger remnant-driven jet}
\begin{figure*}
	\begin{center}
		\includegraphics[width=0.47\textwidth]{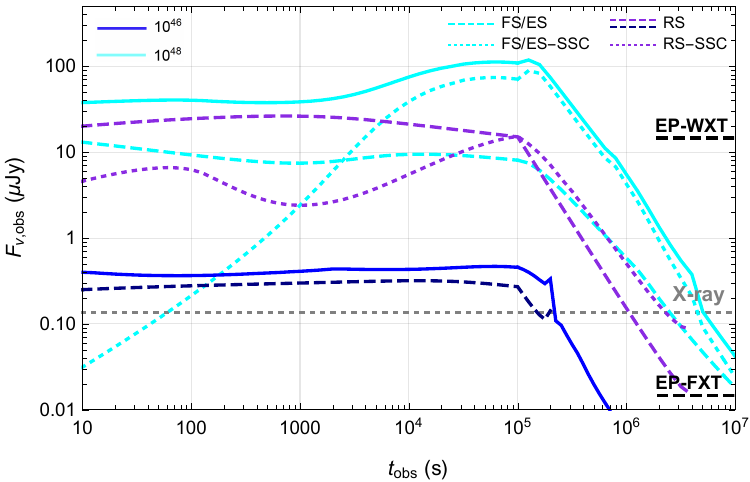}
		\includegraphics[width=0.47\textwidth]{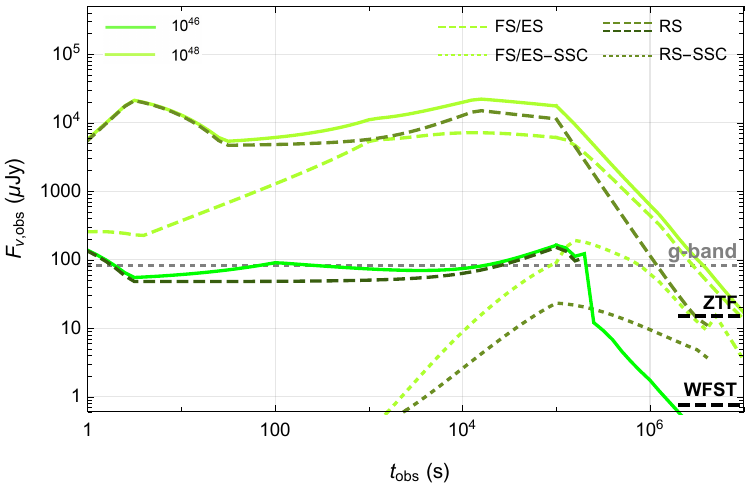}
		\includegraphics[width=0.47\textwidth]{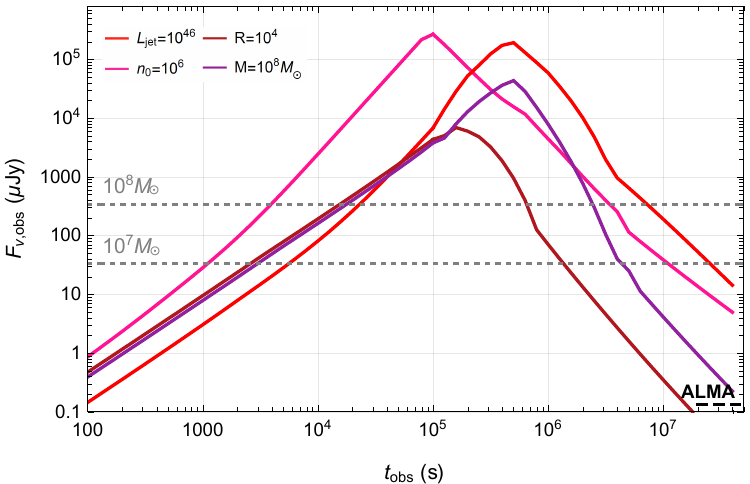}
		\includegraphics[width=0.47\textwidth]{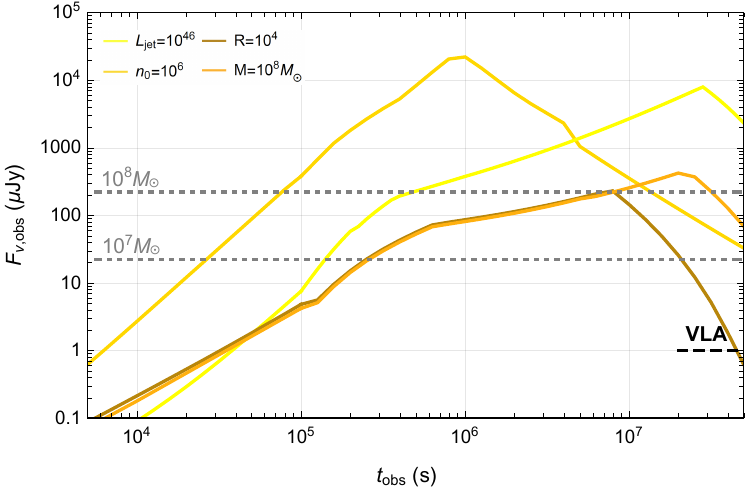}
	\end{center}
	\caption{Multiband light curves of BBH merger remnant-driven
		jet emission. The jet is set as $L_{\text{j}}=10^{46}\ergs$ 
		and $t_{\text{j}}=10^5\s$. In the upper panels, the total 
		emission and its radiative components from a more powerful 
		jet with $L_{\text{j}}=10^{48}\ergs$ are also shown.
		The spikes and sharp declines observed in the light curves 
		arise from the artificial termination of reverse shock evolution, 
		along with its radiative contribution abruptly switched off.
		In the lower panels 
		($1 \text{~mm}$ for the left and $10 \text{~GHz}$
		for the right), three additional cases are presented, 
		$n_0$ decreasing to $10^6\cm^{-3}$, BBH merger occurring 
		at $10^4R_g$ of the AGN disk, and an increase 
		in the SMBH mass to $10^8 M_\odot$, where the jet properties
		are recalculated via updated $\dot{M}_{\text{BHL}}$ and
		$t_{\text{BHL}}$. The grey dashed 
		lines indicate the flux level of AGN background with
		SMBH of $10^7 M_\odot$ and $10^8 M_\odot$.
	    Also shown are the detection sensitivities of the optical 
	    telescope, Wide Field Survey Telescope (WFST), for 
	    $30\s$ of exposure time \citep{Lei23}, the 
	    infrared telescope, Atacama Large Millimeter Array (ALMA), 
	    for $1\text{~ks}$ of exposure time, and the radio 
	    telescope, Very Large Array (VLA), for
	    $1\text{~ks}$ of exposure time, calculated by their 
	    sensitivity calculators.}	
	\label{Fig:Flux-BBH}
\end{figure*}

For jets launched by the recoiling remnant of
a BBH merger through accretion in the AGN disk 
\citep{Chen24}, exploring parameter spaces 
spanning kick velocity of $10^7-10^8\cm\s^{-1}$, 
AGN disk radius of $10^2-10^5R_g$, and central 
SMBH mass of $10^6-10^9M_\odot$, the jet power 
$L_{\text{j}}$ reaches up to $O(10^{46})\ergs$
for $\eta=0.05$, with the case presented in 
Section \ref{LJ} serving as a representative 
example.  

BBH merger remnant-driven jets are inefficient
emitters of X-ray and optical radiation. As
shown in Figure \ref{Fig:Flux-BBH}, for a
$L_{\text{j}}=10^{46}\ergs$ jet, RS 
synchrotron emission produces a detectable 
soft X-ray flare lasting $\sim t_{\text{j}}$;
however, the flare luminosity slightly exceeds 
the AGN background by only a small factor, 
limiting its significance. In the optical band, 
the jet emission—also dominated by RS 
synchrotron radiation—is comparable to 
the AGN background, rendering unambiguous 
identification challenging.  
In contrast, these jets produce prominent 
transient emissions in the infrared and radio 
bands. Whether by reducing the ambient density 
at $10^3R_g$ to $10^6\cm^{-3}$, setting the BBH
merger at $10^4R_g$, or increasing the SMBH mass 
to $10^8 M_\odot$, the jets can produce detectable 
infrared flares persisting for tens to hundreds of days.
The same detectability is also achieved in the 
radio band, where these flares lag behind their
infrared counterparts and persist for extended 
duration. However, radio flares are detectable 
only in AGN hosting low-mass SMBH, otherwise 
the background is intrinsically brighter, while
the jet-induced flare emission in this 
environment is weaker. The light curve properties 
of these flares are generally consistent with 
those generated by GRB jets. Besides, the 
infrared flare displays an additional 
distinctive feature: a steep rise at
$t_\text{obs}\sim 10^5\s$, arising from 
the abrupt enhancement of FS deceleration 
following the end of FS-RS evolution. Most
importantly, the time delay between jet launch 
and the onset of these detectable 
low-frequency transients is less than 
$O(10^5-10^6)\s$,
facilitating their identification as EM 
counterparts to BBH merger GW events.

Increasing the jet power enables its 
X-ray and optical emission to sufficiently 
surpass the AGN background flux,
e.g., to $L_{\text{j}}=10^{48}\ergs$,
as shown in Figure \ref{Fig:Flux-BBH}.
The long-lived RS synchrotron emission 
and the SSC scattering in FS successively 
dominate the X-ray flux, yielding an 
approximately flat and gradually rising 
light curve within 
$\sim t_{\text{j}}=10^5\s$, followed by a 
decay. RS synchrotron emission also 
dominates the early optical flux, whereas 
FS synchrotron emission takes over 
in the late phase, the resulting light
curve exhibits a similar plateau to 
decline structure. In BBH merger
remnant-driven jet scenario, although 
reducing the GW kick velocity could 
boost the jet power as 
$\dot{M}_{\text{BHL}}\propto v_{\text{k}}^{-3}$,
the associated time delay between 
merger and jet launch,
which $\sim t_{\text{BHL}}\propto v_{\text{k}}^{-3}$
\citep{Kaaz23, Chen24}, 
increases concomitantly.
To launch a jet with 
$L_{\text{j}}=3.3\times10^{46}\ergs$, 
the remnant BH of mass $100M_\odot$ moving 
through the AGN disk at $R_0=10^3R_g$ 
from a SMBH of mass $M=10^7M_\odot$
requires a kick velocity
as low as $10^7\cm\s^{-1}$, yielding
$t_{\text{BHL}}=4.4\times10^6\s$.
Given such a prolonged time delay, the 
association between the observed EM 
flares and the GW event cannot be 
robustly established, which severely 
hampers observational follow-up.   
However, the recoil kick of the merger 
remnant can trigger a strong shock 
in the circum-BBH accretion disk, 
enhancing gas angular momentum loss 
and thereby driving enhanced mass 
accretion \citep{Rossi10}; under this
mechanism, a higher efficiency $\eta=1$ 
can substantially amplify the jet power 
to produce identifiable EM counterparts
with short time delays \citep{Tagawa23}.  

In short, BBH merger remnant-driven jets are 
detectable in the infrared and radio 
bands, but remain practically undetectable 
in the X-ray and optical bands unless the 
effective energy conversion efficiency is 
sufficiently high.  

\subsection{Associated thermal emission}
As a jet propagating through the AGN 
accretion disk, a substantial fraction 
of its energy is deposited into the 
cocoon \citep[e.g.,][]{Bromberg11}. 
The subsequent cocoon breakout and 
adiabatic cooling expansion drive 
thermal radiation
\citep{Tagawa24, Rodriguez-Ramirez24, Chen25}, 
accompanying the nonthermal emission 
produced by the jet. 

For powerful uncollimated jets, 
such as successful GRB jets, the 
thermal flare in the soft X-ray band is 
substantially weaker than its nonthermal 
counterpart in both duration ($<O(10^3)\s$) 
and luminosity ($<O(10^{47})\ergs$), 
as shown in Figure 9 of \cite{Chen25}; 
consequently, its contribution to the 
total observed flux is negligible. 
The optical thermal flare emerges at 
$>O(10^4)\s$, by which time the nonthermal 
flare has faded significantly, whereas 
the thermal component
persists for a longer duration
$>O(10^5)\s$, contributing to the observed 
optical radiation. For long-lived
free jets, e.g., BBH merger remnant-driven
jets, the optical thermal flare is dimmer
than the AGN background and therefore
undetectable. In contrast, the cooling 
cocoon produces a bright transient
flare in the soft X-ray band, where 
the jet nonthermal emission is 
intrinsically weak; this thermal 
cocoon component thus serves as
a soft X-ray counterpart to the BBH merger
event. For jets choked within the AGN disk, 
cocoon emission constitutes the sole 
EM signature associated with the 
jet-launching events.

\section{Summary and Discussion}
\label{Sec5}
In this study, motivated by the possibility that relativistic
jets embedded in an AGN accretion disk can successfully
break out, we have investigated their dynamical evolution and
nonthermal emission properties as they propagate through the
AGN environment composed of disk-driven winds.
By analyzing three representative jet systems\textemdash 
namely, a powerful uncollimated jet, a long-lived free jet, 
and an under-accelerated jet at breakout\textemdash we found that the 
jet power and duration jointly govern its global evolution,
while the high-density AGN gas gives rise to two 
characteristic features: rapid dynamical deceleration of the 
jet ejecta, which induces a prompt downshift in the  
SED, and persistent strong synchrotron self-absorption, 
resulting in a distinct quasi-thermal hump in the emission 
spectrum. Successful relativistic jets can produce 
observable EM radiations that substantially 
outshine the AGN background. 
For BNS/NS-BH mergers or massive stellar core collapse, 
the resulting GRB jets produce soft
X-ray and optical flares lasting several to tens of days, 
together with infrared and radio transients persisting for 
months to years. For BBH mergers, jets driven by accretion 
onto the remnant similarly generate long-lived infrared and 
radio transients; detectability in the X-ray and optical 
bands, however, depends strongly on the jet power, making the 
identification of jet‑launching mechanism in BBH mergers 
crucial for predicting their multi‑wavelength signatures. 
Therefore, binary compact object mergers capable of launching 
successful relativistic jets are detectable.
A key advantage is that rapid dissipation of the jet kinetic 
energy leads to prompt multiband emission, thereby shortening
the time delay between the GW trigger and its EM counterpart, 
which greatly facilitates a secure association.

Although our analysis focuses on specific illustrative jet 
systems, the dynamical evolution and the resulting multiband 
emission of any successfully breaking-out jet can be robustly 
predicted using the physical framework developed in this 
study. Combined with our previous work \cite{Chen25}, which 
characterizes the thermal emission from jets breaking out of 
the AGN disk and from the subsequent cocoon cooling expansion, 
we present a comprehensive framework for predicting the EM 
signatures of jets launched in AGN disks \textemdash regardless 
of whether they are eventually choked or successfully emerge.

For the AGN environment, we have adopted a simplified, globally
uniform wind model. However, AGN outflows exhibit rich structural
complexity, spanning a wide range of column densities, 
velocities, and ionization states across distance scales; 
even at a fixed radius, outflow properties vary substantially
among individual AGNs \citep{Laha21}. First, AGN winds are not
continuous or global, but rather sporadic and spatially
fragmented, with different outflow types coexisting in distinct
yet overlapping regions. Consequently, the ambient gas density 
distribution along the jet propagation axis is inherently 
inhomogeneous and multi-phase. Since variations in ambient 
density strongly modulate jet dynamics and the resulting emission 
evolution, the associated multiband light curves may display 
complex and anomalous variability patterns. Second, although 
typical outflow velocities are lower than those of the
propagating jet, extreme cases reach up to $0.2 c$, comparable 
to or even exceeding the velocity of jet ejecta during the deep 
Newtonian phase. Such high-velocity outflows can substantially 
alter jet dynamics and the corresponding radiative signatures. 
Third, because the ambient gas is only partially ionized,
photoionization induced by the jet radiation field would 
raise time- and frequency-dependent opacity, leading to distinctive
spectral evolution \citep{Ray23}. In summary, this study outlines the 
general characteristics of jet evolution and associated its radiation 
under idealized environmental assumptions. To interpret specific 
observed events, detailed modeling of multiband light curves will 
require a tailored construction of the host AGN environment.

Additionally, we have assumed a uniform top-hat jet structure and an
on-axis observer geometry. In reality, however, jet material undergoing
breakout naturally develops an angular distribution in quantities 
such as LF and energy density, indicating that successful AGN-disk 
breakout jets are intrinsically structured \citep[e.g.,][]{Gottlieb21}. 
Since both the jet structure and the observer viewing angle strongly 
affect the emission detectability and variability 
properties \citep[e.g.,][]{Kathirgamaraju24, Pang24}, the radiative 
evolution of off-axis structured jets requires further systematic 
investigation. Furthermore, while a median AGN SED is adopted to 
estimate the detectability of jet emission, the dispersion among the 
AGN SED population is large \citep[e.g.,][]{Shang11}. Therefore, 
for robust modeling of any specific jet event, the individual SED 
of the host AGN must be employed as the background radiation field.

\section{Acknowledgments}
We would like to thank the referee for valuable comments and Sen-Lin Pang for
helpful discussions. This work was supported by the National Natural Science Foundation of China (grant No. 12393812), and the Strategic Priority Research Program of the Chinese Academy of Sciences (grant NO. XDB0550300).

\begin{appendix}
\section{Synchrotron self-absorption frequency}
\label{va}
In the strong self-absorption regime, where $\nu'_a >\nu'_c$, 
the thermal hump dominates over the initial synchrotron
emission around $\nu'_a$, thereby self-consistently 
setting the absorption frequency \citep{Kobayashi04}. 
Under a heating-cooling balance, 
the total energy of self-absorbed synchrotron radiation $E_{a}$ 
is reprocessed and emitted around $\nu'_a$ in a dynamical time, 
and the corresponding specific emission luminosity can be 
estimated as 
$L_{\nu'_a}\simeq E_{a}/\nu'_a t'$.

When the spectral ordering satisfies 
$\nu'_a >\nu'_m>\nu'_c$, the initial 
self-absorbed synchrotron photons 
consume the energy deposited in electrons 
between $\gamma_m$ and $\gamma_a$ following 
the distribution of Equation \eqref{eqN}, i.e.,
$E_{a}\sim \frac{p-1}{p-2} \xi_{e} N_\text{tot} 
\gamma_m m_e c^2/(1+Y_{\text{TOT}})$,
and thereby
\begin{equation}
L_{\nu'_a}= C \left(\frac{\nu'_m \nu'_c}{\nu'^{2}_a}\right)^{\frac{1}{2}}
\frac{\bar{\gamma}_{c}}{\gamma_c} L_{\nu \text {,max}},
\label{Lva}
\end{equation}
with $C= \frac{16\pi}{27\sqrt{3}} \frac{p-1}{p-2}$, 
$L_{\nu \text {,max }} = \xi_{e} N_\text{tot} 
P_{\nu\text {,max}}$, 
and the factor $\bar{\gamma}_{c}/\gamma_c$ arises 
because only a fraction of the total relativistic 
electrons contribute to the instantaneous emission 
on account of the very fast cooling of electron when 
$\bar{\gamma}_{c}\ll 1$ \citep{Beniamini13, Rahaman25}.
This luminosity exceeds the unabsorbed synchrotron 
contribution at $\nu'_a$, which reads
\begin{equation}
L_{\nu'_a}= C \left(\frac{\nu'_m}{\nu'_a}\right)^{1-\frac{p}{2}} L_{\nu'_{a},\text{syn}},
\end{equation}
where \citep{Zhang18}
\begin{equation}
L_{\nu'_{a},\text{syn}}=\left(\frac{\nu'_c}{\nu'_m}\right)^{\frac{1}{2}}
\left(\frac{\nu'_m}{\nu'_a}\right)^{\frac{p}{2}} 
\frac{\bar{\gamma}_{c}}{\gamma_c} L_{\nu \text {,max }}.
\end{equation}
Consequently, a distinctive spectral hump appears 
around $\nu'_a$, above which the spectrum resumes the
standard synchrotron power-law form 
\citep{Ghisellini88}. When
$\nu'_m>\nu'_a >\nu'_c$, the majority of photon 
energy is successfully radiated around $\nu'_m$, 
while the self-absorbed energy below $\nu'_a$ is 
reduced by a factor of $(\nu'_a/\nu'_m)^{1/2}$
relative to the $\nu'_a >\nu'_m>\nu'_c$ case, 
and correspondingly
\begin{equation}
L_{\nu'_a}= C \left(\frac{\nu'_c}{\nu'_a}\right)^{\frac{1}{2}}
\frac{\bar{\gamma}_{c}}{\gamma_c} L_{\nu \text {,max }}
=  C L_{\nu'_{a},\text{syn}},
\end{equation}
which is slightly larger than the original 
synchrotron emission. For the case of 
$\nu'_a >\nu'_c>\nu'_m$, only electrons
with LF larger than $\gamma_c$ cool 
effectively on a dynamical time. The 
self-absorption energy now corresponds
to the energy deposited in electrons between 
$\gamma_c$ and $\gamma_a$, i.e.,
$E_a \sim \frac{p-1}{p-2}
\left(\gamma_c/\gamma_m\right)^{1-p} \gamma_c N_\text{tot} 
m_e c^2/(1+Y_{\text{TOT}}) $, and thus
\begin{equation}\label{Lva2}
L_{\nu'_a}=C \left(\frac{\nu'_c}{\nu'_m}\right)^{\frac{1-p}{2}}
\frac{\nu'_c}{\nu'_a} L_{\nu \text {,max }}
=  C \left(\frac{\nu'_c}{\nu'_a}\right)^{1-\frac{p}{2}} L_{\nu'_{a},\text{syn}},
\end{equation}
where \citep{Zhang18}
\begin{equation}
L_{\nu'_{a},\text{syn}}=\left(\frac{\nu'_c}{\nu'_m}\right)^{-\frac{p-1}{2}}
\left(\frac{\nu'_c}{\nu'_a}\right)^{\frac{p}{2}} L_{\nu \text {,max }}.
\end{equation}

Employing
$I_{\nu'_a}=2\gamma_a m_e \nu'^2_a$ and estimating 
$I_{\nu'_a}\sim L_{\nu'_{a}}/4\pi^2\theta^2r^2$, the synchrotron 
self-absorption frequency is derived as   
\begin{equation}
\nu'_a=\left(\frac{C I_{\nu, \max }}{2 m_{e}} \sqrt{\frac{3 e B'}{2 \pi m_{e} c}}\right)^{\frac{2}{7}}
\left\{
\begin{array}{lr}
\left(\nu'_m \nu'_c\right)^{\frac{1}{7}}, 
&\nu'_c< \nu'_m< \nu'_a \\[1ex]
\nu'^{\frac{1}{7}}_c, 
& \nu'_c< \nu'_a< \nu'_m \\[1ex]
\nu'^{\frac{2}{7}}_c  \left(\frac{\nu'_c}{\nu'_m}\right)^{\frac{1-p}{7}}, 
&\nu'_m< \nu'_c< \nu'_a \\[1ex]
\end{array}\right.
\end{equation}
where 
$I_{\nu, \max } \sim \xi_{e} (\bar{\gamma}_{c}/\gamma_{c})
N_\text{tot} P_{\nu\text {,max }} /4\pi^2\theta^2r^2$.

In the weak self-absorption regime where $\nu'_a <\nu'_c$,
employing Equation \eqref{vaT}, the synchrotron 
self-absorption frequency is given by \citep{Wang26}
\begin{equation}
\nu'_a=\left\{
\begin{array}{lr}\left(\frac{I_{\nu, \max }}{2\gamma_{p} m_{e} \nu'^{1 / 3}_p}\right)^{\frac{3}{5}}, 
& \nu'_{a}<\min \left[\nu'_{m}, \nu'_{c}\right] \\[2.5ex]
\left(\frac{I_{\nu, \max }}{2 m_{e}} \sqrt{\frac{3 e B'}{2 \pi m_{e} c}}\right)^{\frac{2}{p+4}} \nu'^{\frac{p-1}{p+4}}_{m}, 
& \nu'_{m}<\nu'_{a}<\nu'_{c} \end{array}\right.
\end{equation}
where $\nu'_p = \min \left[\nu'_{m}, \nu'_{c}\right]$.

\section{Electron energy distributions and synchrotron radiation spectra}
\label{FN}
\setcounter{equation}{0}
In the strong absorption regime, where $\nu_a> \nu_c$, 
synchrotron self-absorption thermalizes the electron 
population below $\gamma_a$, driving both the electron 
energy distribution and the emitted synchrotron spectrum 
toward a quasi-thermal form. This results in a 
thermal bump in the electron distribution near $\gamma_a$,
accompanied by a corresponding pronounced thermal 
hump at $\nu_a$. Above $\gamma_a$ and  $\nu_a$, 
the distributions recover their standard 
power-law forms \citep{Ghisellini88}.

For the case $\gamma_a>\gamma_m>\gamma_c$, 
the heated electrons with LF around 
$\gamma_a$ share $E_a$, the
corresponding number density can be estimated as 
$N_{e}(\gamma_a) \sim (\gamma_m/\gamma_a) \xi_e N_{\text{tot}}$.
Therefore, the modified energy distribution of the 
instantaneously radiating electrons is 
\begin{equation}\label{Nem}
\frac{d N_{e}}{d \gamma}=N_{e,\text{syn}}
\begin{cases}
3 \gamma_{m} \gamma^{-4}_{a} \gamma^{2}, 
& \gamma<\gamma_{a} \\[1.5ex]
\gamma_{m}^{p-1} \gamma_{c} \gamma^{-p-1}, 
& \gamma>\gamma_{a}
\end{cases}
\end{equation}
where $N_{e,\text{syn}}= \xi_{e} (\bar{\gamma}_{c}/\gamma_{c})
N_\text{tot}$. Here, the low-$\gamma$ branch 
represents the quasi-thermal excess induced 
by SSA, while the high-$\gamma$ branch 
retains the standard cooled power law.
Adopting the peak specific luminosity from
Equation \eqref{Lva}, the observed synchrotron 
spectrum is approximately given by
\begin{equation}\label{Fm}
F_{\nu,\text{syn}}(\nu) = F_\text{syn,max} e^{-\frac{\nu}{\nu_M}}
\begin{cases}
C \left(\frac{\nu_m \nu_c}{\nu^{2}_a}\right)^{\frac{1}{2}} 
\left(\frac{\nu}{\nu_a}\right)^2, 
& \nu<\nu_{a} \\[2.5ex]
\left(\frac{\nu_c}{\nu_m }\right)^{\frac{1}{2}} 
\left(\frac{\nu_m}{\nu_a }\right)^{\frac{p}{2}} 
\left(\frac{\nu}{\nu_a}\right)^{-\frac{p}{2}}, 
& \nu>\nu_{a}
\end{cases}
\end{equation}
with an exponential cutoff at high energies,  
where $F_\text{syn,max}=
\Gamma N_{e,\text{syn}} P_{\nu\text {,max }} 
/4\pi D^2_{\text{L}}$ for an on-axis observer, and 
$D_{\text{L}}$ is the luminosity distance of the 
emission source. In reality, the actual spectral 
transition across the thermal hump near $\nu_a$ 
is smoothed rather than a sharp cutoff; 
nevertheless, the piecewise power-law approximation
given by Equation \eqref{Fm} robustly captures 
the dominant spectral shape \citep{Gao13a}.
To ensure spectral continuity, we 
model the emission above $\nu_a$ using a 
simplified quasi-thermal spectrum \citep{Kobayashi04}, 
where the flux declines beyond the peak as
$\exp (1-\nu/\nu_a)$, until connecting  
to the non-thermal part of the spectrum.

For the case $\gamma_m>\gamma_a>\gamma_c$, 
most of the electrons are cooling to 
$\gamma_a$, and the modified energy 
distribution is
\begin{equation}
\frac{d N_{e}}{d \gamma}=N_{e,\text{syn}}
\begin{cases}
3 \gamma^{-3}_{a}\gamma^{2}, 
& \gamma<\gamma_{a} \\[1ex]
\gamma_{c} \gamma^{-2}, 
& \gamma_{a}<\gamma<\gamma_{m} \\[1ex]
\gamma_{m}^{p-1} \gamma_{c} \gamma^{-p-1}. 
& \gamma>\gamma_{m}
\end{cases}
\end{equation} 
The synchrotron spectrum is approximately 
given by
\begin{equation}
F_{\nu,\text{syn}}(\nu) = F_\text{syn,max} e^{-\frac{\nu}{\nu_M}}
\begin{cases}
C \left(\frac{\nu_c}{\nu_a }\right)^{\frac{1}{2}} 
\left(\frac{\nu}{\nu_a}\right)^2, 
& \nu<\nu_{a} \\[2.5ex]
\left(\frac{\nu_c}{\nu_a }\right)^{\frac{1}{2}} 
\left(\frac{\nu}{\nu_a}\right)^{-\frac{1}{2}}, 
& \nu_{a}<\nu<\nu_{m} \\[2.5ex]
\left(\frac{\nu_c}{\nu_m }\right)^{\frac{1}{2}} 
\left(\frac{\nu}{\nu_m}\right)^{-\frac{p}{2}}. 
& \nu>\nu_{m}
\end{cases}
\end{equation}

For the case $\gamma_a>\gamma_c>\gamma_m$, 
the electron number density around $\gamma_a$
can be estimated as 
$N_{e,\text{SSA}} \sim E_a/\gamma_a m_e c^2$,
and 
\begin{equation}
\frac{d N_{e}}{d \gamma}=N_{e,\text{syn}}
\begin{cases}
3 \gamma_{m}^{p-1} \gamma_{c}^{2-p} \gamma^{-4}_{a} \gamma^{2}, 
& \gamma<\gamma_{a} \\[1.5ex]
(p-1)\gamma_{m}^{p-1} \gamma_{c} \gamma^{-p-1}.
& \gamma>\gamma_{a}
\end{cases}
\end{equation}
Adopting the peak specific luminosity of
Equation \eqref{Lva2}, the synchrotron spectrum 
is thus given by
\begin{equation}
F_{\nu,\text{syn}}(\nu) = F_\text{syn,max} e^{-\frac{\nu}{\nu_M}}
\begin{cases}
C \left(\frac{\nu_c}{\nu_a}\right)
\left(\frac{\nu_c}{\nu_m }\right)^{\frac{1-p}{2}} 
\left(\frac{\nu}{\nu_a}\right)^2, 
& \nu<\nu_{a} \\[2.5ex]
\left(\frac{\nu_c}{\nu_m }\right)^{\frac{1-p}{2}} 
\left(\frac{\nu_c}{\nu_a }\right)^{\frac{p}{2}} 
\left(\frac{\nu}{\nu_a}\right)^{-\frac{p}{2}}.
& \nu>\nu_{a}
\end{cases}
\end{equation}

In the weak self-absorption regime, where $\nu_a < \nu_c$,
the electron energy distribution and the 
corresponding synchrotron spectrum have been 
extensively studied and are now well established 
within the standard afterglow model  
\citep{Sari98, Granot02, Gao13b, Wang26}.
We therefore adopt this standard framework 
without further elaboration. 

\end{appendix}

\end{document}